\documentclass[english,twocolumn,superscriptaddress]{revtex4-2}
\usepackage{amsmath}

\usepackage{newtxmath}
\usepackage[T1]{fontenc}
\usepackage[latin9]{inputenc}
\usepackage{graphicx}

\makeatletter
\usepackage{newtxtext}
\usepackage{newtxmath}
\usepackage{bbm}
\usepackage{hyperref}
\usepackage{xcolor}
\definecolor{BSHGreen}{RGB}{0, 100, 0}
\hypersetup{
    colorlinks=true,
    linkcolor=BSHGreen,
    citecolor=BSHGreen,
    urlcolor=BSHGreen
}
\allowdisplaybreaks[4]
\usepackage{slashed}

\renewcommand\theequation{\arabic{equation}}

\makeatother

\usepackage{babel}
\begin{document}
\title{Suppression of Extrinsic Anomalous Hall Conductivity in Disordered
Parity Anomalous Semimetal}
\author{Shi-Hao Bi}
\address{Department of Physics, The University of Hong Kong, Pokfulam Road,
Hong Kong, China}
\author{Bo Fu}
\address{School of Sciences, Great Bay University, Dongguan 523000, China}
\author{Shun-Qing Shen}
\email{sshen@hku.hk}

\address{Department of Physics and State Key Laboratory of Optical Quantum
Matter, The University of Hong Kong, Pokfulam Road, Hong Kong, China}
\date{\today}
\begin{abstract}
We present an analytical investigation of the extrinsic contributions
to the anomalous Hall conductivity in the context of the half-quantized
Hall effect observed in disordered parity anomalous semimetal emerged
from semi-magnetic topological insulator thin films. The gapless Dirac
cone surface state, which embodies the quintessence of the half-quantized
Hall effect, exhibits remarkable robustness against disorder scattering.
Two primary extrinsic mechanisms, the side-jump and skew-scattering,
are deemed irrelevant and make no contributions. These results establish
the parity anomalous semimetal as a disorder-resilient quantum phase,
thereby providing insights into Dirac fermion physics.
\end{abstract}
\maketitle

\section{Introduction}

The quantum anomalous Hall effect is characterized by the quantization
of Hall conductivity in the absence of an external magnetic field
\citep{Haldane1988:PRL,yu2010quantized,chu2011surface,qiao2014quantum,ChangCZ2023:RMP,Shen2017:TI}.
This phenomenon arises from the non-trivial Berry curvature of occupied
Bloch bands, resulting in the emergence of an integer-quantized Hall
plateau within the band gap. The Hall conductivity can be represented
by an integer in units of $e^{2}/h$, with the integer corresponding
to the Chern number, a topological invariant of the occupied bands
\citep{TKNN1982:PRL,xiao2010berry}. Owing to this topological protection,
the quantum anomalous Hall insulator exhibits remarkable robustness
against impurity scattering \citep{Prodan2011:JPA}. Moreover, the
quantization of the Hall conductivity can survive despite the presence
of strong disorder whose magnitude vastly exceeds the bulk band gap.
Experimentally, the quantum anomalous Hall effects have been observed
in various materials, such as Cr/V-doped topological insulator thin
film of $\left(\mathrm{Bi},\text{Se}\right)_{2}\mathrm{Te}_{3}$ family
\citep{ChangCZ2013:SCI,checkelsky2014trajectory,kou2014scale,ChangCZ2015:NM},
the intrinsic magnetic topological insulator $\mathrm{MnBi_{2}Te_{4}}$
\citep{Otrokov2017:2DMat,Otrokov2019:PRL,Otrokov2019:Nat,Gong2019:CPL,Li2019:SciAdv,ZhangYB2020:SCI,FX2023:NSR,Li2023:NSR},
and the moir{\'e} materials of graphene \citep{Goldhaber-Gordon2019:SCI,Chen2020:Nat,AFYoung2020:SCI}
or transition-metal dichalcogenides \citep{MakKF2021:Nat,XXD2023:SCI}.

In recent years, the notion of topological invariants has been extended
to gapless quantum systems, leading to the discovery of quantum anomalous
semimetals \citep{Shen2022:npjQM}. The quantum anomalous semimetals
constitute a family of exotic quantum states of matter that incorporate
the quantum anomaly of gapless Dirac fermions in high-energy physics
with solid-state systems, featuring half-quantized topological invariants
\citep{Shen2022:npjQM,WangHW2022:PRB,Zou2022:PRB,Zou2023:PRB,Shen2024:Coshare,Fu2024:NC,WangHW2024:PRB,WangHW2025:PRB,FuBo2025:CommPhys,Bi2025:CommPhys,Bi2025:PRB}.
In two-dimensional systems, a prototypical example is the parity anomalous
semimetal (PAS) \citep{Zou2022:PRB,Zou2023:PRB}, which displays explicit
symmetry breaking at high energy states in the first Brillouin zone
that violates either parity or time-reversal symmetry, while exhibiting
a half-quantized Hall conductivity. Experimentally, such half-quantized
Hall conductivity have been reported in semi-magnetic topological
insulators \citep{Mogi2022:NatPhys} and in tri-layer magnetic topological
insulator heterostructure \citep{Wang-prl2026,Zhuo-prl-2026,Yang-AdvMater-2026}.
Zou \textit{et al}. developed an analytical framework, which revealed
the pivotal contribution of the gapless Dirac cone to this exotic
phase \citep{Zou2022:PRB,Zou2023:PRB}. The longitudinal conductivity
has a minimal value when the chemical potential lies near the gapless
Dirac point, which was also measured recently \citep{Wang-prl2026,fu2025nearly}.
Owing to its metallic character, the persistence of half-quantized
Hall conductivity in PAS under strong disorder conditions remains
a pertinent question. Notwithstanding some numerical studies that
have corroborated its robustness \citep{Bi2025:CommPhys,Bi2025:PRB},
a comprehensive understanding of this phenomenon demands further investigation.

\begin{figure}
\centering\includegraphics[width=8.5cm]{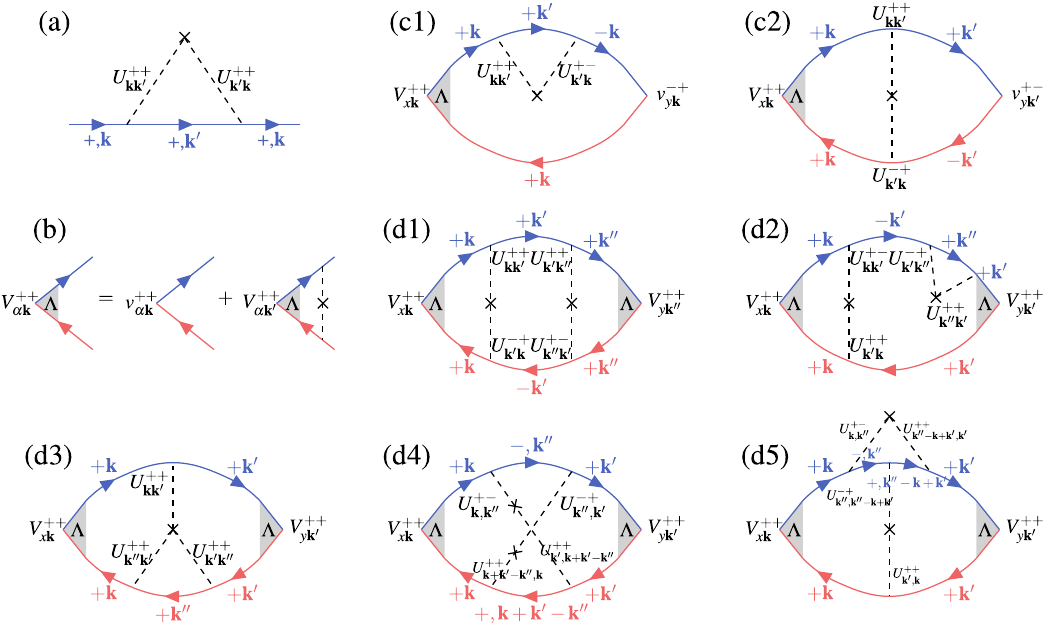}

\caption{Feynman diagrams for the extrinsic mechanisms of anomalous Hall conductivity.
(a) The self-energy within the self-consistent Born approximation.
(b) Ladder diagrams for the disorder-renormalized velocity operator
$V_{\alpha{\bf k}}^{++}$. The shaded area, denoted by $\Lambda$,
is the diffuson contribution to the vertex correction of the bare
velocity $v_{\alpha{\bf k}}^{++}$. (c1, c2) The diagrammatic representation
of side-jump mechanism for the anomalous Hall conductivity. The diagrammatic
representation of the (d1, d2) intrinsic and (d3) extrinsic skew scattering
mechanism, together with the (d4) X and (d5) $\Psi$ diagram for coherent
skew scattering. Each diagram represents a unique family of contributions,
which always appear in complex conjugate pairs. $U_{{\bf kk}'}$ denotes
the scattering matrix element, and the dashed lines represent the
disorder-averaged correlation functions of them. The left-bound (blue)
and right-bound (red) arrow lines correspond to the retarded ($G^{R})$
and advanced ($G^{A}$) Green's functions, respectively.}

\label{fig:AHE}
\end{figure}

In metallic systems with spin-orbital coupling, impurity scatterings
usually lead to contribution to anomalous Hall conductivity, known
as side-jump mechanism and skew-scattering \citep{SMIT195839,BERGER19641141,Sinitsyn2007:PRB,NPOng2010:RMP,LuHZ2013:PRB}.
In the semiclassical wave-packet theory, disorder scattering induces
a transverse coordinate shift, which gives rise to both a side-jump
velocity and an anomalous correction to the distribution function
\citep{Sinitsyn2007:PRB}. Collectively, these contributions constitute
the side-jump mechanism. In parallel, an asymmetric scattering rate
leads to the skew-scattering. A systematic comparisons between the
kinetic Boltzmann approach and the Kubo-Streda formalism have facilitated
the development of Feynman diagrammatic techniques to study these
mechanisms \citep{Sinitsyn2007:PRB}, as depicted in Fig. \ref{fig:AHE}.
Therefore, it is an essential question whether these extrinsic mechanisms
can destroy the quantization of the Hall conductivity in PAS or semi-magnetic
topological insulator thin film, or whether they are suppressed by
the peculiar structure of the PAS.

In this work, we address this question by performing a thorough investigation
of the extrinsic mechanisms contributing to the anomalous Hall effect
in a disordered PAS, encompassing both non-magnetic and magnetic Anderson
impurities. Within the framework of self-consistent Born approximation
and Kubo formalism, we demonstrate that the disorder scattering will
not lead to a gap opening for the massless Dirac cone and that the
standard side-jump and skew-scattering mechanisms make no net contribution
to the half-quantized Hall conductance. Thus the PAS retains its half-quantized
Hall response even when extrinsic contributions from disorder are
taken into account.

The remainder of this paper is organized as follows. In Sec. \ref{sec:Model},
we introduce the model for PAS and discuss the correlation functions
of disorder scattering matrices. Then, in Sec. \ref{sec:AHE}, we
compute the extrinsic Hall conductivity arising from side-jump and
skew-scattering mechanisms, using the Feynman diagrams shown in Fig.
\ref{fig:AHE}. Finally, we summarize our findings and provide a discussion
in Sec. \ref{sec:Conclusion}.

\section{Model of Parity Anomalous Semimetal\label{sec:Model}}

\begin{figure}
\centering\includegraphics[width=7.5cm]{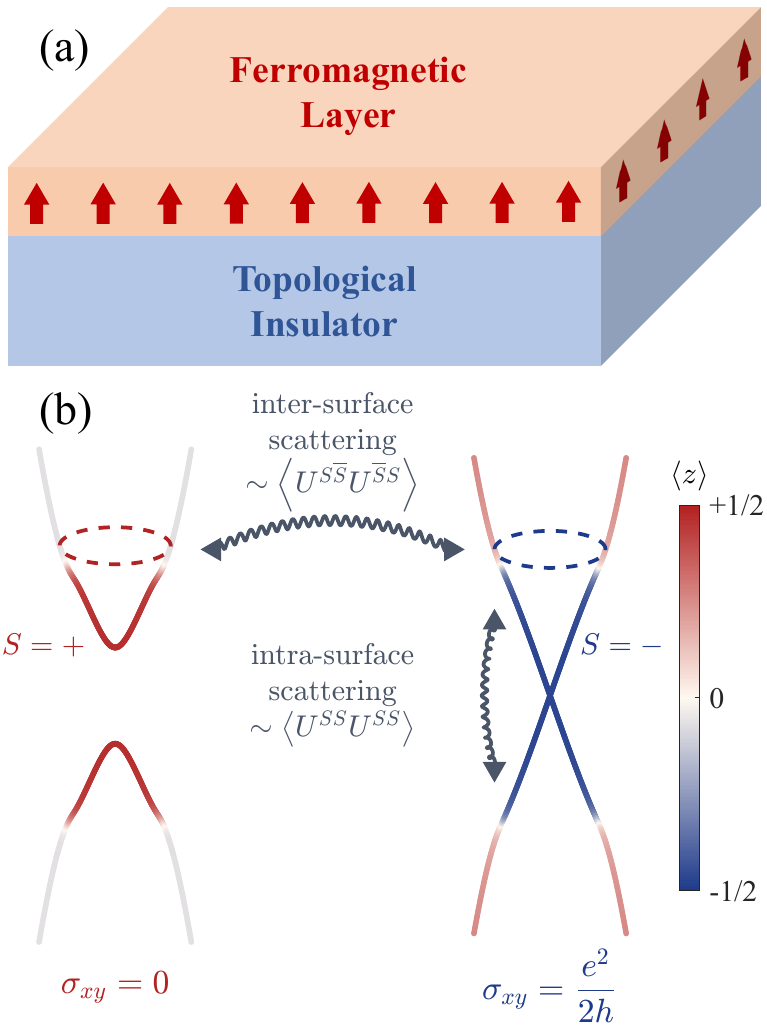}\caption{(a) Schematic illustration of a semi-magnetic topological insulator
film. (b) Intra- and inter-surface scattering in gapless and gapped
Dirac cone states. The color denotes the average $z$-position distribution.}

\label{fig:model}
\end{figure}

In a semi-magnetic structure comprising thin films of three-dimensional
topological insulators, the PAS phase can be realized. This phenomenon
emerges from the regulation of a single gapless Dirac cone on the
two-dimensional surface, which is notable for its half-quantized Hall
conductivity. The pristine system is a $\left(\mathrm{Bi},\text{Se}\right)_{2}\mathrm{Te}_{3}$
thin film, which is a strong topological insulator with a bulk energy
gap of about $0.28$ eV \citep{Zhang2009:NatPhys}. Notably, this
material hosts a single gapless Dirac cone on both the top and bottom
surfaces. First-principles calculations suggest that the topological
nature of this compound can be effectively captured by a tight-binding
model constructed from the four orbitals near the Fermi level, $|P1_{-}^{+},\pm\frac{1}{2}\rangle$
and $|P2_{+}^{-},\pm\frac{1}{2}\rangle$, which predominantly originate
from $\left(\mathrm{Bi},\text{Se}\right)$ and $\mathrm{Te}$ atoms.
The corresponding lattice Hamiltonian is 
\begin{equation}
H=\sum_{{\bf r}_{i}}\Psi_{{\bf r}_{i}}^{\dagger}M_{0}\Psi_{{\bf r}_{i}}+\sum_{{\bf r}_{i},\alpha=x,y,z}\Psi_{{\bf r}_{i}}^{\dagger}\mathcal{T}_{\alpha}\Psi_{{\bf r}_{i}+{\bf e}_{\alpha}}+\mathrm{H.c.},\label{eq:H-1}
\end{equation}
where $M_{0}=\left(m_{0}-4t_{\parallel}-2t_{\perp}\right)\sigma_{0}\tau_{z}$,
and $\mathcal{T}_{\alpha}=t_{\alpha}\sigma_{0}\tau_{z}-\frac{\mathrm{i}\lambda_{\alpha}}{2}\sigma_{\alpha}\tau_{x}$.
In-plane isotropy implies $t_{x}=t_{y}=t_{\parallel}$ and $\lambda_{x}=\lambda_{y}=\lambda_{\parallel}$.
$\Psi_{{\bf r}_{i}}^{\dagger}$ and $\Psi_{{\bf r}_{i}}$ are the
electron creation and annihilation operators at site ${\bf r}_{i}$.
$\sigma_{\alpha}$ and $\tau_{\alpha}$ are Pauli matrices acting
on the spin and orbital spaces, respectively. If the magnetic elements
such as $\mathrm{Cr}$ are doped on the top $n_{z}$ layers and form
a ferromagnetic order, a Zeeman field term is added to the Hamiltonian:
\begin{equation}
H_{\mathrm{Z}}=\sum_{i_{z}<n_{z}}\Psi_{{\bf r}_{i}}^{\dagger}V_{0}\sigma_{z}\tau_{0}\Psi_{{\bf r}_{i}},\label{eq:ZM}
\end{equation}
which opens a gap of magnitude $2V_{0}$ for the top surface Dirac
cone. Nevertheless, the Dirac cone on the bottom surface remains gapless
and contributes a quantum Hall conductivity of $\frac{e^{2}}{2h}$
\citep{Zou2023:PRB}. See Fig. \ref{fig:model} for an illustration.

Unlike the intrinsic Hall conductivity, which arises from the Berry
curvature of the Bloch bands, the extrinsic Hall conductivity is generated
by impurity scattering on the Fermi surface. To elucidate the influence
of disorder scattering on the Hall conductivity, we explicitly include
both non-magnetic and magnetic impurities in our analysis. The impurity
scattering is modeled by the on-site random potential 
\begin{equation}
U({\bf r})=\sum_{i,\alpha}u_{\alpha,i}\sigma_{\alpha}\tau_{0}\delta_{{\bf r},{\bf r}_{i}},\label{eq:Uimp}
\end{equation}
where the random amplitudes $u_{\alpha,i}$ are uniformly distributed
within the range $\left[-W_{\alpha}/2,+W_{\alpha}/2\right]$, where
$\alpha$ iterates over $0,x,y$, and $z$. Here, $\sigma_{0}$ is
the $2\times2$ unit matrix that represents the non-magnetic (spin-independent)
scatterings, whereas the remaining Pauli matrices in spin space account
for the random magnetic scatterings. Without loss of generality, we
make the simplifying assumption that distinct types of scattering
mechanisms are statistically independent. Under this assumption, the
disorder-averaged correlations of the random potentials are 
\begin{equation}
\left\langle u_{\alpha,i}u_{\beta,j}\right\rangle =\frac{W_{\alpha}^{2}}{12}\delta_{\alpha\beta}\delta_{ij},\label{eq:2nd-moment}
\end{equation}
where $\left\langle \cdots\right\rangle $ denotes the disorder-averaging.
With this microscopic disorder model in place, we are now equipped
to investigate the extrinsic mechanisms that contribute to the anomalous
Hall conductivity in the following section.

\subsection{Mathematical Preliminaries}

To gain deeper insight into the half-quantized Hall effect, we start
from the surface subband description of the thin-film topological
insulator. The Hamiltonian governing both the gapless and massive
Dirac cone surface states has been derived analytically in Refs. \citep{Zou2023:PRB,KZ2024:SciPost}.
We briefly summarize the main steps and notation used in this work.
First, we impose periodic boundary conditions in the $x$ and $y$
directions and apply a Fourier transformation to derive the Hamiltonian
$H\left({\bf k}\right)$ in momentum space. This yields an expression
of the form:
\begin{equation}
H\left({\bf k}\right)=\sum_{i_{z},{\bf k}}\Psi_{i_{z},{\bf k}}^{\dag}\mathcal{E}_{i_{z}}\Psi_{i_{z},{\bf k}}+\Psi_{i_{z},{\bf k}}^{\dag}\mathcal{T}_{z}\Psi_{i_{z}+1,{\bf k}}+\mathrm{H.c.},\label{eq:H-2}
\end{equation}
where $\mathcal{E}_{i_{z}}=\sum_{\alpha}\lambda_{\parallel}\sin k_{\alpha}\sigma_{\alpha}\tau_{x}+\left[m_{0}({\bf k})-2t_{\perp}\right]\sigma_{0}\tau_{z}$,
$\alpha$ runs over $x$ and $y$, and 
\begin{equation}
m_{0}({\bf k})=m_{0}-4t_{\parallel}\left(\sin^{2}\frac{k_{x}}{2}+\sin^{2}\frac{k_{y}}{2}\right).
\end{equation}
The analytical derivation of the Hamiltonian of the surface Dirac
cone states for this lattice model has been established in Refs. \citep{Shan2010:NJP,Zou2023:PRB,KZ2023:PRB}.
We adopt the same framework and define the four basis states as follows:
\begin{equation}
\begin{array}{cc}
\left|\Phi_{i_{z},{\bf k}}^{+,\uparrow}\right\rangle =\left[\begin{array}{c}
\varphi_{{\bf k}}\left(+,l_{z}\right)\\
0
\end{array}\right], & \left|\Phi_{i_{z},{\bf k}}^{+,\downarrow}\right\rangle =\left[\begin{array}{c}
0\\
\chi_{{\bf k}}\left(-,l_{z}\right)
\end{array}\right],\vspace{0.5em}\\
\left|\Phi_{i_{z},{\bf k}}^{-,\uparrow}\right\rangle =\left[\begin{array}{c}
\chi_{{\bf k}}\left(+,l_{z}\right)\\
0
\end{array}\right], & \left|\Phi_{i_{z},{\bf k}}^{-,\downarrow}\right\rangle =\left[\begin{array}{c}
0\\
\varphi_{{\bf k}}\left(-,l_{z}\right)
\end{array}\right].
\end{array}\label{eq:basis-1}
\end{equation}
where the spinors $\varphi_{{\bf k}}$ and $\chi_{{\bf k}}$ are defined
as
\begin{equation}
\begin{array}{c}
\varphi_{{\bf k}}\left(s,l_{z}\right)=C_{{\bf k}}\left[\begin{array}{c}
-\mathrm{i}s\lambda_{\perp}f_{{\bf k},+}\left(l_{z}\right)\\
t_{\perp}\eta_{1}f_{{\bf k},-}\left(l_{z}\right)
\end{array}\right],\vspace{0.5em}\\
\chi_{{\bf k}}\left(s,l_{z}\right)=C_{{\bf k}}\left[\begin{array}{c}
t_{\perp}\eta_{1}f_{{\bf k},-}\left(l_{z}\right)\\
\mathrm{i}s\lambda_{\perp}f_{{\bf k},+}\left(l_{z}\right)
\end{array}\right],
\end{array}\label{eq:basis-2}
\end{equation}
and $l_{z}=i_{z}-\frac{L_{z}+1}{2}$ is layer index measured from
the center of the film. Here, $C_{{\bf k}}$ is a normalization constant,
and the explicit forms of $\eta_{{\bf k},1}$ and $f_{{\bf k},\pm}$
are (for $l=\frac{L_{z}+1}{2}$) \citep{Shan2010:NJP,Zou2023:PRB,KZ2023:PRB}
\begin{equation}
\begin{aligned}\eta_{{\bf k},1}= & \frac{-2\left(\cos\xi_{1}-\cos\xi_{2}\right)}{\sin\xi_{1}\cot\xi_{1}l-\sin\xi_{2}\cot\xi_{2}l},\\
f_{{\bf k},+}= & \frac{\cos\xi_{1}l_{z}}{\cos\xi_{1}l}-\frac{\cos\xi_{2}l_{z}}{\cos\xi_{2}l},\\
f_{{\bf k},-}= & \frac{\sin\xi_{1}l_{z}}{\sin\xi_{1}l}-\frac{\sin\xi_{2}l_{z}}{\sin\xi_{2}l},
\end{aligned}
\label{eq:basis-3}
\end{equation}
where $\xi_{1,2}$ obey the transcendental equations \citep{Shan2010:NJP,Zou2023:PRB,KZ2023:PRB}
\begin{equation}
\begin{aligned}m= & M+2t_{\perp}\frac{\cos\xi_{1}g\left(\xi_{1}\right)-\cos\xi_{2}g\left(\xi_{2}\right)}{g\left(\xi_{1}\right)-g\left(\xi_{2}\right)},\\
\cos\xi_{\alpha}= & \frac{-Mt_{\perp}-\left(-1\right)^{\alpha}\sqrt{M^{2}t_{\perp}^{2}-\Omega\left(M^{2}+\lambda_{\perp}^{2}-m^{2}\right)}}{2\Omega},
\end{aligned}
\end{equation}
where $\Omega=t_{\perp}^{2}-\frac{\lambda_{\perp}^{2}}{4}$, $M=m_{0}({\bf k})-2t_{\perp}$
and $g\left(\xi\right)=\frac{\tan\xi l}{\sin\xi}$.

In above representation, the surface Dirac cone Hamiltonian reads
\begin{equation}
H_{0}({\bf k})=\lambda_{\parallel}\left(\sin k_{x}\sigma_{x}+\sin k_{y}\sigma_{y}\right)+m({\bf k})\tau_{z}\sigma_{z},\label{eq:H-3}
\end{equation}
where $m({\bf k})=m_{0}({\bf k})\Theta\left[-m_{0}({\bf k})\right]$
describes how the gapless Dirac cone surface states seamlessly merge
into the bulk ($\Theta$ is the Heaviside step function). For the
Zeeman term, its projection onto the surface states subspace takes
the form of 
\[
H_{\mathrm{Z}}({\bf k})=\frac{V_{0}f_{1}({\bf k})}{2}\left(\tau_{0}-\tau_{y}\right)\sigma_{z},
\]
where $f_{1}({\bf k})=\Theta\left[m_{0}({\bf k})\right]+\alpha\Theta\left[-m_{0}({\bf k})\right]$
with $\alpha\simeq n_{z}/L_{z}$ parameterizing the effect of energy
split at high energy regime. The total clean Hamiltonian in the surface
subspace then reads 
\begin{equation}
H({\bf k})=H_{0}({\bf k})+H_{\mathrm{Z}}({\bf k}).\label{eq:H-4}
\end{equation}
To disentangle the massive and gapless surface sectors, we apply a
unitary transformation $\mathcal{U}=\exp\left(-\frac{\mathrm{i}}{2}\vartheta_{{\bf k}}\tau_{x}\sigma_{0}\right)$
which brings the Hamiltonian into block-diagonal form: 
\begin{equation}
H_{S}=\mathcal{U}H\mathcal{U}^{-1}=\lambda_{\parallel}\left(\sin k_{x}\sigma_{x}+\sin k_{y}\sigma_{y}\right)+\mathcal{M}_{S}({\bf k})\sigma_{z},\label{eq:H-5}
\end{equation}
with $\vartheta_{{\bf k}}=\arctan\frac{V_{0}f_{1}({\bf k})}{2m({\bf k})}$
and 
\begin{equation}
\mathcal{M}_{\pm}({\bf k})=\frac{V_{0}f_{1}({\bf k})}{2}\pm\sqrt{m({\bf k})^{2}+\left[\frac{V_{0}f_{1}({\bf k})}{2}\right]^{2}}.
\end{equation}
The $S=+$ block corresponds to the massive Dirac cone state, where
$\mathcal{M}_{+}({\bf k})$ is positive in the whole first Brillouin
zone, and thus it is topologically trivial. In contrast, the $S=-$
block describes the gapless Dirac cone; here, the high energy states
explicitly breaks the time-reversal symmetry and generates a half-quantized
Hall conductivity.

In the calculation of the Feynman diagram, we adopt the continuum
limit of the Hamiltonian given above:
\begin{equation}
H_{S}\left({\bf k}\right)=\lambda_{\parallel}\left(k_{x}\sigma_{x}+k_{y}\sigma_{y}\right)+\mathcal{M}_{S}({\bf k})\sigma_{z}.\label{eq:H-6}
\end{equation}
 For $S=-$, we approximate $\mathcal{M}_{S}({\bf k})$ as ($k=\left|{\bf k}\right|$
denotes the magnitude of the momentum ${\bf k}$)
\begin{equation}
\mathcal{M}_{-}({\bf k})\approx\begin{cases}
0 & ,k<k_{c}\\
m_{0}-t_{\parallel}k^{2} & ,k\geqslant k_{c}
\end{cases}.
\end{equation}
Here, $k_{c}\approx\sqrt{m_{0}/t_{\parallel}}$ represents a critical
momentum demarcating the boundary between the gapless low-energy regime
and the massive high-energy regime within the first Brillouin zone.
In the low-energy regime ($k<k_{c}$), the mass term disappears, rendering
the Hamiltonian symmetric with respect to both parity and time-reversal
operations. As a consequence, the Berry curvature is constrained such
that its integral along any constant-energy contour is zero, even
though the local Berry curvature can be nonzero due to, for example,
hexagonal warping effects \citep{FuLiang2009:PRL,LiuCX2010:PRB,LiuCX2013:PRL}.
In the high-energy regime ($k\geqslant k_{c}$), the finite mass term
breaks time-reversal symmetry, allowing for a net Berry curvature
on equal-energy lines and thereby giving rise to the half-quantized
Hall response \citep{Zou2023:PRB,FuBo2025:CommPhys}. On the other
hand, for $S=+$, we similarly approximate $\mathcal{M}_{S}({\bf k})$
as 
\begin{equation}
\mathcal{M}_{+}({\bf k})\approx\begin{cases}
V_{0} & ,k<k_{c}\\
t_{\parallel}k^{2}-m_{0} & ,k\geqslant k_{c}
\end{cases}.
\end{equation}
Notably, the mass term remains consistently positive throughout the
first Brillouin zone, thereby rendering the corresponding band topologically
trivial with zero Hall conductivity.

\subsection{Correlation Functions of Scattering Amplitudes}

Upon making the unitary transformation, the basis states in Eq. (\ref{eq:basis-1})
are transformed into $\left|\Psi_{i_{z},{\bf k}}^{S,s}\right\rangle =\mathcal{U}_{{\bf k}}^{SR;sr}\left|\Phi_{i_{z},{\bf k}}^{R,r}\right\rangle $.
In this representation, the scattering amplitude between momenta ${\bf k}$
and ${\bf k}'$ for an impurity potential of the $\alpha$-type takes
the form of (with ${\rm S}$ the sample area) 
\begin{equation}
U_{\alpha,{\bf k}{\bf k}'}^{SS';ss'}=\frac{1}{{\rm S}}\sum_{i_{z},{\bf r}_{j}}u_{\alpha,i_{z},{\bf r}_{j}}\mathrm{e}^{-\mathrm{i}\left({\bf k}-{\bf k}'\right)\cdot{\bf r}_{j}}\left\langle \Psi_{i_{z},{\bf k}}^{S,s}\right|\sigma_{\alpha}\tau_{0}\left|\Psi_{i_{z},{\bf k}'}^{S',s'}\right\rangle .\label{eq:SA-1}
\end{equation}
Note that the indices $S$ (surface block) and $s$ (spin) are decoupled,
enabling us to factorize the unitary transformation as $\mathcal{U}_{{\bf k}}^{SR;sr}=\mathcal{U}_{{\bf k}}^{SR}\sigma_{0}^{sr}$.
Therefore, the scattering amplitude Eq. (\ref{eq:SA-1}) simplifies
to 
\begin{equation}
\begin{aligned}U_{\alpha,{\bf k}{\bf k}'}^{SS';ss'}= & \frac{1}{{\rm S}}\sum_{i_{z},{\bf r}_{j}}u_{\alpha,i_{z},{\bf r}_{j}}\mathrm{e}^{-\mathrm{i}\left({\bf k}-{\bf k}'\right)\cdot{\bf r}_{j}}\left(\mathcal{U}_{{\bf k}}^{SR}\right)^{*}\mathcal{U}_{{\bf k}'}^{S'R'}\\
 & \times\left\langle \Phi_{i_{z},{\bf k}}^{R,s}\right|\sigma_{\alpha}\tau_{0}\left|\Phi_{i_{z},{\bf k}'}^{R',s'}\right\rangle .
\end{aligned}
\label{eq:SA-2}
\end{equation}
Moreover, the spin indices can be separated from the expectation value
in Eq. (\ref{eq:SA-2}) as $\left\langle \Phi_{i_{z},{\bf k}}^{R,s}\right|\sigma_{\alpha}\tau_{0}\left|\Phi_{i_{z},{\bf k}'}^{R',s'}\right\rangle \sim\sigma_{\alpha}^{ss'}$,
so that all remaining structure resides in the surface subspace ($R,R'$)
and the layer index ($i_{z}$).

\begin{figure}
\centering\includegraphics[width=8.5cm]{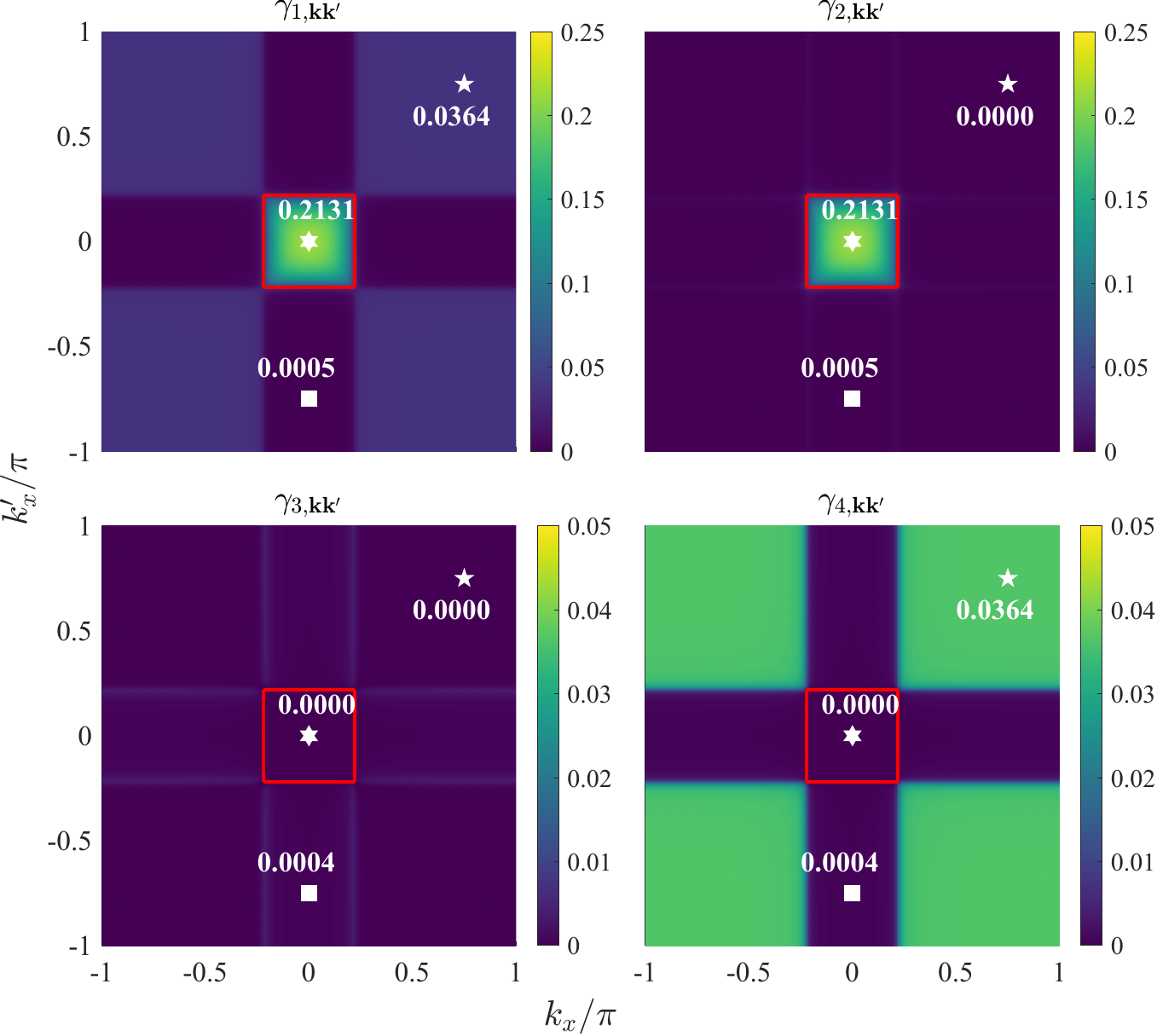}

\caption{Numerical calculation of $\ensuremath{\gamma_{i,{\bf k}{\bf k}'}}$.
Typical values are labeled near the markers. We set $k_{y}=k_{y}^{\prime}=0$,
and $L_{z}=40$. The red boxes indicate the parity-invariant regime.}
\label{fig:correlation}
\end{figure}

With the scattering amplitudes determined, we now compute their correlation
functions. By employing the disorder correlator defined in Eq. (\ref{eq:2nd-moment})
, the disorder-averaged two-amplitude correlation function is given
by: 
\begin{equation}
\begin{aligned}\left\langle U_{\alpha,{\bf k}{\bf k}'}^{SS';ss'}U_{\alpha,{\bf k}'{\bf k}}^{PP';pp'}\right\rangle  & =\frac{W_{\alpha}^{2}}{12{\rm S}}\left(\mathcal{U}_{{\bf k}}^{SR}\right)^{*}\mathcal{U}_{{\bf k}'}^{S'R'}\left(\mathcal{U}_{{\bf k}'}^{PQ}\right)^{*}\mathcal{U}_{{\bf k}}^{P'Q'}\\
\times\sum_{i_{z}}\left\langle \Phi_{i_{z},{\bf k}}^{R,s}\right| & \sigma_{\alpha}\tau_{0}\left|\Phi_{i_{z},{\bf k}'}^{R',s'}\right\rangle \left\langle \Phi_{i_{z},{\bf k}'}^{Q,p}\right|\sigma_{\alpha}\tau_{0}\left|\Phi_{i_{z},{\bf k}}^{Q',p'}\right\rangle .
\end{aligned}
\label{eq:SA-4}
\end{equation}
After some algebra, the correlation functions of the scattering matrix
with two amplitudes scattering matrix can be cast in the compact form
\begin{subequations}
\renewcommand{\theequation}{\theparentequation\alph{equation}} %
\begin{align}
\left\langle U_{0/z,{\bf k}{\bf k}'}^{SS}U_{0/z,{\bf k}'{\bf k}}^{SS}\right\rangle =&\frac{W_{0/z}^{2}}{12{\rm S}}\mathcal{F}_{1,{\bf k}{\bf k}'}\sigma_{0/z}\otimes\sigma_{0/z}, \\
\left\langle U_{0/z,{\bf k}{\bf k}'}^{S\overline{S}}U_{0/z,{\bf k}'{\bf k}}^{\overline{S}S}\right\rangle =&\frac{W_{0/z}^{2}}{12{\rm S}}\mathcal{F}_{2,{\bf k}{\bf k}'}\sigma_{0/z}\otimes\sigma_{0/z}, \\
\left\langle U_{x/y,{\bf k}{\bf k}'}^{SS}U_{x/y,{\bf k}'{\bf k}}^{SS}\right\rangle =&\frac{W_{x/y}^{2}}{12{\rm S}}\mathcal{F}_{3,{\bf k}{\bf k}'}\sigma_{x/y}\otimes\sigma_{x/y}, \\
\left\langle U_{x/y,{\bf k}{\bf k}'}^{S\overline{S}}U_{x/y,{\bf k}'{\bf k}}^{\overline{S}S}\right\rangle =&\frac{W_{x/y}^{2}}{12{\rm S}}\mathcal{F}_{4,{\bf k}{\bf k}'}\sigma_{x/y}\otimes\sigma_{x/y},
\end{align}
\label{eq:SA-5}
\end{subequations}where the momentum-dependent prefactors are\begin{subequations}
\label{eq:SA-6}
\renewcommand{\theequation}{\theparentequation\alph{equation}} %
\begin{align}
\mathcal{F}_{1,{\bf k}{\bf k}'}=&\cos^{2}{\displaystyle \frac{\vartheta_{{\bf k}}-\vartheta_{{\bf k}'}}{2}}\gamma_{1,{\bf k}{\bf k}'}+\sin^{2}{\displaystyle \frac{\vartheta_{{\bf k}}+\vartheta_{{\bf k}'}}{2}}\gamma_{2,{\bf k}{\bf k}'},\\
\mathcal{F}_{2,{\bf k}{\bf k}'}=&\sin^{2}{\displaystyle \frac{\vartheta_{{\bf k}}-\vartheta_{{\bf k}'}}{2}}\gamma_{1,{\bf k}{\bf k}'}+\cos^{2}{\displaystyle \frac{\vartheta_{{\bf k}}+\vartheta_{{\bf k}'}}{2}}\gamma_{2,{\bf k}{\bf k}'},\\
\mathcal{F}_{3,{\bf k}{\bf k}'}=&\cos^{2}{\displaystyle \frac{\vartheta_{{\bf k}}+\vartheta_{{\bf k}'}}{2}}\gamma_{3,{\bf k}{\bf k}'}+\sin^{2}{\displaystyle \frac{\vartheta_{{\bf k}}-\vartheta_{{\bf k}'}}{2}}\gamma_{4,{\bf k}{\bf k}'},\\
\mathcal{F}_{4,{\bf k}{\bf k}'}=&\sin^{2}{\displaystyle \frac{\vartheta_{{\bf k}}+\vartheta_{{\bf k}'}}{2}}\gamma_{3,{\bf k}{\bf k}'}+\cos^{2}{\displaystyle \frac{\vartheta_{{\bf k}}-\vartheta_{{\bf k}'}}{2}}\gamma_{4,{\bf k}{\bf k}'}.
\end{align}
\end{subequations}The quantities $\gamma_{i,\mathbf{k}\mathbf{k}'}$ are wave-function
shape factors defined by \begin{subequations}
\label{eq:SA-7}
\renewcommand{\theequation}{\theparentequation\alph{equation}} %
\begin{align}
\gamma_{1,{\bf k}{\bf k}'}=&\left|C_{{\bf k}}C_{{\bf k}'}\right|^{2}{\displaystyle \sum_{i_{z}}}\left|\lambda_{\perp}^{2}f_{{\bf k},+}^{*}f_{{\bf k}',+}+t_{\perp}^{2}\eta_{{\bf k},1}\eta_{{\bf k}',1}f_{{\bf k},-}^{*}f_{{\bf k}',-}\right|^{2},\\
\gamma_{2,{\bf k}{\bf k}'}=&\left|C_{{\bf k}}C_{{\bf k}'}\right|^{2}{\displaystyle \sum_{i_{z}}}\lambda_{\perp}^{2}t_{\perp}^{2}\left|f_{{\bf k},+}^{*}\eta_{{\bf k}',1}f_{{\bf k}',-}+\eta_{{\bf k},1}f_{{\bf k},-}^{*}f_{{\bf k}',+}\right|^{2},\\
\gamma_{3,{\bf k}{\bf k}'}=&\left|C_{{\bf k}}C_{{\bf k}'}\right|^{2}{\displaystyle \sum_{i_{z}}}\lambda_{\perp}^{2}t_{\perp}^{2}\left|f_{{\bf k},+}^{*}\eta_{{\bf k}',1}f_{{\bf k}',-}-\eta_{{\bf k},1}f_{{\bf k},-}^{*}f_{{\bf k}',+}\right|^{2},\\
\gamma_{4,{\bf k}{\bf k}'}=&\left|C_{{\bf k}}C_{{\bf k}'}\right|^{2}{\displaystyle \sum_{i_{z}}}\left|\lambda_{\perp}^{2}f_{{\bf k},+}^{*}f_{{\bf k}',+}-t_{\perp}^{2}\eta_{{\bf k},1}\eta_{{\bf k}',1}f_{{\bf k},-}^{*}f_{{\bf k}',-}\right|^{2}.
\end{align}
\end{subequations}For $k<k_{c}$, we have $\vartheta_{{\bf k}}=\pi/2$; for $k>k_{c}$,
$\vartheta_{{\bf k}}\approx0$. The momentum dependence of the $\gamma_{i}$
is shown in Fig. \ref{fig:correlation}, which will be used to analyze
the self-energy correction in the following section.

In the the skew scattering diagram, the three-amplitude correlation
function $\left\langle U_{\alpha,{\bf kk}'}^{SS}U_{\alpha,{\bf k}'{\bf k}''}^{SS}U_{\alpha,{\bf k}''{\bf k}}^{SS}\right\rangle $
appears. However, such contributions vanish identically because
\begin{equation}
\left\langle u_{\alpha,i_{z},{\bf r}_{j}}u_{\alpha,i_{z}^{\prime},{\bf r}_{j}^{\prime}}u_{\alpha,i_{z}^{\prime\prime},{\bf r}_{j}^{\prime\prime}}\right\rangle =0.
\end{equation}
 In the subsequent analysis of extrinsic anomalous Hall conductivity,
our primary focus will be on the gapless Dirac cone state, where the
$S=-$ superscript is omitted. The correlation function of the intra-
and inter-band scattering amplitudes, for instance, can be obtained
as 
\begin{equation}
\left\langle U_{0,{\bf k}{\bf k}'}^{++}U_{0,{\bf k}'{\bf k}}^{+-}\right\rangle =\frac{W_{0}^{2}}{12{\rm S}}\ensuremath{\mathcal{F}_{1,{\bf k}{\bf k}'}}\left\langle +{\bf k}\right|\sigma_{0}\left|+{\bf k}'\right\rangle \left\langle +{\bf k}'\right|\sigma_{0}\left|-{\bf k}\right\rangle .
\end{equation}
where $\left|\pm{\bf k}\right\rangle $ are the Bloch states for conduction/valence
band of the effective surface Dirac Hamiltonian.

\section{Extrinsic Mechanism of Anomalous Hall Conductivity\label{sec:AHE}}

The disorder scattering induces a self-energy correction to the unperturbed
Hamiltonian, which can be separated into two components: a Hermitian
component that can be absorbed into the dispersion relation, and an
imaginary component that represents the finite lifetime $\tau$ of
quasi-particle excitation near the Fermi surface. A critical concern
is whether magnetic impurity scattering will destroy the gapless Dirac
cone state, potentially opening an energy gap at the $\Gamma$ point.
As a consequence of this energy gap opening, the erstwhile gapless
Dirac cone state undergoes a transition into a massive one, thereby
compromising the underlying mechanism of the half-quantized Hall effect.

To reveal the stability of the gapless Dirac cone, we employ the self-consistent
Born approximation to compute the self-energy \citep{Groth2009:PRL},
a methodology that remains valid in the weak scattering regime with
$\epsilon_{F}\tau\gg1$: 
\begin{equation}
\Sigma_{S}\left(\epsilon_{F},{\bf k}\right)=\sum_{{\bf k}',R}\left\langle U_{{\bf k}{\bf k}'}^{SR}\frac{1}{\epsilon_{F}-H_{R}({\bf k}')-\Sigma_{R}+\mathrm{i}0^{+}}U_{{\bf k}'{\bf k}}^{RS}\right\rangle ,\label{eq:SCBA-1}
\end{equation}
where $S,R=\pm$ label the two surface subband sectors. The disorder-induced
energy gap at the $\Gamma$ point in the gapless Dirac cone can be
extracted as $\Delta_{\Gamma}=\frac{1}{2}\mathrm{Tr}\left[\Sigma_{-}\left(0,{\bf 0}\right)\sigma_{z}\right]$.

The momentum integral can be partitioned into two regimes: low-energy
and high-energy parts, respectively. In the low-energy regime where
$k'<k_{c}$, the relevant terms that could potentially open a gap
in the $S=-$ sector are
\begin{equation}
\Sigma_{-}\left(0,{\bf k};k'<k_{c}\right)=\int_{k'<k_{c}}\frac{\mathrm{d}^{2}{\bf k}'}{\left(2\pi\right)^{2}}\sum_{\alpha=x,y}\mathcal{K}_{-}^{\alpha}({\bf k}'),
\end{equation}
with the integrand denoted by
\begin{equation}
\mathcal{K}_{S}^{\alpha}({\bf k}')=\frac{W_{0/z}^{2}}{12}\sigma_{\alpha}{\displaystyle \frac{\gamma_{1,{\bf kk}'}+\gamma_{2,{\bf kk}'}}{\mathrm{i}0^{+}-H_{S}({\bf k}')-\Sigma_{S}}}\sigma_{\alpha}
\end{equation}
for $\alpha=0,z$ and 
\begin{equation}
\mathcal{K}_{S}^{\alpha}({\bf k}')=\frac{W_{x/y}^{2}}{12}\sigma_{\alpha}{\displaystyle \frac{\gamma_{3,{\bf kk}'}+\gamma_{4,{\bf kk}'}}{\mathrm{i}0^{+}-H_{S}({\bf k}')-\Sigma_{S}}}\sigma_{\alpha}
\end{equation}
for $\alpha=x,y$. Nevertheless, one can show that for $k,k'<k_{c}$,
the wavefunction shape factors $\gamma_{3,{\bf kk}'}$ and $\gamma_{4,{\bf kk}'}$
vanish (see Appendix \ref{app:B} for a detailed proof). This is also
confirmed numerically in Fig. \ref{fig:correlation}. Therefore, we
have $\Sigma_{-}\left(0,{\bf 0};k'<k_{c}\right)=0$ explicitly and
there is no low-energy contribution to $\Delta_{\Gamma}$. In the
high-energy regime where $k'>k_{c}$, the relevant contributions arise
from 
\begin{equation}
\Sigma_{-}\left(0,{\bf k};k'>k_{c}\right)=\frac{1}{2}\int_{k'>k_{c}}\frac{\mathrm{d}^{2}{\bf k}'}{\left(2\pi\right)^{2}}\sum_{R,\alpha}\mathcal{K}_{R}^{\alpha}({\bf k}').
\end{equation}
As the high-energy mass terms in $H_{+}$ and $H_{-}$ possess opposite
signs, the intra- and inter-subband scatterings are mutually canceled:
\begin{equation}
\sum_{R}\sigma_{\alpha}\frac{1}{\mathrm{i}0^{+}-H_{R}({\bf k}')-\Sigma_{R}}\sigma_{\alpha}=0.\label{eq:cancel}
\end{equation}
In conclusion, our thorough examination unequivocally confirms that
the gapless Dirac cone exhibits remarkable robustness against disorder
scattering even when magnetic impurities are doped.

Before proceeding, we briefly compare the above results with the two-dimensional
Wilson fermion model \citep{Wilson1974:PRD,Shen2022:npjQM,Zou2023:PRB},
\begin{equation}
H_{{\rm W}}\left({\bf k}\right)=\lambda_{\parallel}\left(k_{x}\sigma_{x}+k_{y}\sigma_{y}\right)-t_{\parallel}\left(k_{x}^{2}+k_{y}^{2}\right)\sigma_{z}.
\end{equation}
This model exhibits a half-quantized Hall conductance at the Dirac
point, given by $\sigma_{xy}\left(\epsilon_{F}=0\right)=-\frac{e^{2}}{2h}{\rm sgn}t_{\parallel}$.
In the presence of the short-range Anderson impurity potential given
by Eq. (\ref{eq:Uimp}) (ignoring $\tau_{0}$ component), the retarded
self-energy at $\epsilon_{F}=0$ reduces to
\begin{equation}
\Sigma_{{\rm W}}^{R}=\sum_{\alpha}\frac{W_{\alpha}^{2}}{12}\int\frac{{\rm d}^{2}{\bf k}}{\left(2\pi\right)^{2}}\sigma_{\alpha}\frac{1}{{\rm i}0^{+}-H_{{\rm W}}\left({\bf k}\right)-\Sigma^{R}}\sigma_{\alpha}.\label{eq:SCBA-2}
\end{equation}
Assuming the decomposition
\begin{equation}
\Sigma_{{\rm W}}^{R}=-{\rm i}\kappa\sigma_{0}+m\sigma_{z},
\end{equation}
and substitute into Eq. (\ref{eq:SCBA-2}), we obtain the self-consistent
equations\begin{subequations}
\label{eq:SCBA-3}
\begin{align}
\kappa=&w_{1}\int\frac{{\rm d}^{2}{\bf k}}{\left(2\pi\right)^{2}}\frac{\kappa}{\kappa^{2}+\lambda_{\parallel}^{2}k^{2}+\left(m-t_{\parallel}k^{2}\right)^{2}},\\m=&w_{2}\int\frac{{\rm d}^{2}{\bf k}}{\left(2\pi\right)^{2}}\frac{t_{\parallel}k^{2}-m}{\kappa^{2}+\lambda_{\parallel}^{2}k^{2}+\left(m-t_{\parallel}k^{2}\right)^{2}}.
\end{align}
\end{subequations}Here, $w_{1}=\sum_{\alpha}W_{\alpha}^{2}/12$ and $w_{2}=\left(W_{0}^{2}+W_{z}^{2}-W_{x}^{2}-W_{y}^{2}\right)/12$.
This indicates that the spin-flip impurities ($\alpha=x,y$) make
opposite contributions to the mass correction compared to spin-conserving
impurities ($\alpha=0,z$). Solving these equations yielding the following
expression:
\begin{equation}
m\approx\frac{w_{2}}{8\pi t_{\parallel}}\ln\frac{t_{\parallel}^{2}\Lambda^{4}}{\kappa^{2}+m^{2}},
\end{equation}
where $\Lambda$ is the ultraviolet cutoff for the momentum integral.
The non-zero solution for $\kappa$ exists provided that $w_{1}\geqslant8m\left|t_{\parallel}\right|$.
Consequently, the Hall conductance at weak scattering regime is corrected
to 
\begin{equation}
\sigma_{xy}\left(\epsilon_{F}=0\right)=-\frac{e^{2}}{2h}\left({\rm sgn}t_{\parallel}+{\rm sgn}m\right),
\end{equation}
which is always an integer multiple of $e^{2}/h$. The Wilson fermion
model acts as the critical point of a topological phase transition
between a normal insulator and a quantum anomalous Hall insulator.
In contrast, for the PAS realized in the semi-magnetic structure of
topological insulator films, the impurity scattering correlation functions
in Eq. (\ref{eq:SA-5}), (\ref{eq:SA-6}), and (\ref{eq:SA-7}) become
highly local in ${\bf k}$-space and decay rapidly outside the parity-invariant
regime, as shown in Fig. (\ref{fig:correlation}) (purple region).
This behavior strongly suppresses momentum integrals for $k>k_{c}$.
Moreover, the cancellation in Eq. (\ref{eq:cancel}) eliminates the
dominant mass correction, thereby protecting the gapless nature of
the PAS.

As the gapless Dirac cone in PAS exhibits unwavering resilience against
disorder scattering, the Fermi surface endures unbroken and persistent.
We will focus on the gapless Dirac cone surface states and the $S=-$
superscript is left out. The arrow lines in Fig. \ref{fig:AHE} represent
the disorder-averaged retarded ($R$) and advanced ($A$) Green's
functions: 
\begin{equation}
G_{s,{\bf k}}^{R/A}=\frac{1}{\epsilon_{F}+s\epsilon_{{\bf k}}\pm\dfrac{\mathrm{i}}{2\tau}}.
\end{equation}
Here $s=+(-)$ is the index for conduction (valence) band, while $\epsilon_{{\bf k}}=\sqrt{\lambda_{\parallel}^{2}k^{2}+\mathcal{M}_{-}^{2}}$
describes the band dispersion. In what follows, a physical quantity
$O$ evaluated at the Fermi energy will be denoted as $O_{F}$. The
total scattering time can be obtained through the imaginary part of
the SCBA self-energy: 
\begin{equation}
\frac{1}{\tau}=2\pi\sum_{{\bf k}'}\left\langle U_{{\bf kk}'}^{++}U_{{\bf k}'{\bf k}}^{++}\right\rangle \delta\left(\epsilon_{F}-\epsilon_{{\bf k}'}\right),
\end{equation}
where the superscripts $\pm$ denote conduction and valence band indices,
respectively. The total scattering rate is given by the sum of individual
contributions from each scattering mechanism $\alpha$, denoted by
$\tau^{-1}=\sum_{\alpha}\tau_{\alpha}^{-1}$. With the disorder-average
correlation function of scattering amplitudes listed in Eq. (\ref{eq:SA-5}),
we find that only the nonmagnetic and $z$-polarized channels contribute:
\begin{equation}
\frac{1}{\tau_{0/z}}=2\pi\rho_{F}\frac{W_{0/z}^{2}\left(\gamma_{1,F}+\gamma_{2,F}\right)}{12}\times\left(a^{4}+b^{4}\right),
\end{equation}
where $a=\cos\frac{\theta_{F}}{2}$, $b=\sin\frac{\theta_{F}}{2}$,
and $\cos\theta_{F}=\mathcal{M}_{-,F}/\epsilon_{F}$.

\subsection{Vertex Correction to the Velocity Operator}

In the Feynman diagrams presented in Fig. \ref{fig:AHE}, the shaded
gray region accounts for the contribution of vertex corrections to
the velocity operator originating from multiple disorder scatterings.
The disorder-renormalized velocity operator $V_{\alpha{\bf k}}^{++}$
can be evaluated as \citep{LuHZ2013:PRB} 
\begin{equation}
V_{\alpha{\bf k}}^{++}=v_{\alpha{\bf k}}^{++}+\sum_{{\bf k}'}v_{\alpha{\bf k}'}^{++}\left\langle U_{{\bf kk}'}^{++}U_{{\bf k}'{\bf k}}^{++}\right\rangle G_{+{\bf k}'}^{R}G_{+{\bf k}'}^{A},
\end{equation}
where $v_{\alpha{\bf k}}^{ss'}=\lambda_{\parallel}\left\langle s{\bf k}\right|\sigma_{\alpha}\left|s'{\bf k}\right\rangle $
denotes the bare velocity operator in the gapless regime, the central
focus of the present work, and the Bloch states $\left|\pm{\bf k}\right\rangle $
for the conduction/valence bands are 
\begin{equation}
\left|+{\bf k}\right\rangle =\left[\begin{array}{c}
a\\
b\mathrm{e}^{\mathrm{i}\phi}
\end{array}\right],\,\left|-{\bf k}\right\rangle =\left[\begin{array}{c}
b\\
-a\mathrm{e}^{\mathrm{i}\phi}
\end{array}\right],
\end{equation}
with $\phi=\arctan\frac{k_{y}}{k_{x}}$. Upon adopting the ansatz
$V_{\alpha{\bf k}}^{++}=\eta_{v}v_{\alpha{\bf k}}^{++}$, we obtain
an expression for the vertex correction coefficient as 
\begin{equation}
\eta_{v}=\left[1-\left(\frac{\tau}{\tau_{0}}-\frac{\tau}{\tau_{z}}\right)\frac{a^{2}b^{2}}{a^{4}+b^{4}}\right]^{-1}.
\end{equation}

\subsection{Side-Jump Mechanism}

We now proceed to evaluate the extrinsic anomalous Hall conductivity
using the Feynman diagrams presented in Fig. \ref{fig:AHE}. We start
from the side-jump mechanism, which describes a transverse shift of
a wave packet upon scattering from spin-orbit-coupled impurities \citep{SMIT195839,BERGER19641141}.
In metallic ferromagnets with spin-orbit coupling, this contribution
is of the same order of magnitude as the intrinsic contribution arising
from the Berry curvature. By permuting the position of the correlation
functions of the scattering amplitudes, we can generate a family of
diagrams, which can be categorized into complex conjugate pairs. It
is sufficient to consider the representative diagrams illustrated
in Fig. \ref{fig:AHE} (c1,c2). They can be calculated as 
\begin{equation}
\sigma_{xy}^{\mathrm{sj}1}=\frac{e^{2}}{h}\int\frac{\mathrm{d}^{2}{\bf k}}{(2\pi)^{2}}\sum_{{\bf k}'}V_{x{\bf k}}^{++}v_{y{\bf k}}^{-+}\left\langle U_{{\bf k}{\bf k}'}^{++}U_{{\bf k}'{\bf k}}^{+-}\right\rangle G_{+{\bf k}}^{R}G_{-{\bf k}}^{R}G_{+{\bf k}}^{A}G_{+{\bf k}'}^{R},
\end{equation}
\begin{equation}
\sigma_{xy}^{\mathrm{sj}2}=\frac{e^{2}}{h}\int\frac{\mathrm{d}^{2}{\bf k}}{(2\pi)^{2}}\sum_{{\bf k}'}V_{x{\bf k}}^{++}v_{y{\bf k}'}^{+-}\left\langle U_{{\bf k}{\bf k}'}^{++}U_{{\bf k}'{\bf k}}^{-+}\right\rangle G_{+{\bf k}}^{R}G_{+{\bf k}}^{A}G_{+{\bf k}'}^{R}G_{-{\bf k}'}^{A}.
\end{equation}
On the Fermi surface, the bare velocity operators are 
\begin{equation}
\left(v_{x{\bf k}}^{++},v_{y{\bf k}}^{++}\right)=\lambda_{\parallel}\sin\theta_{F}\left(\cos\phi,\sin\phi\right),
\end{equation}
\begin{equation}
v_{y{\bf k}}^{+-}=\left(v_{y{\bf k}}^{-+}\right)^{*}=\mathrm{i}\lambda_{\parallel}\left(\cos^{2}\frac{\theta_{F}}{2}\mathrm{e}^{\mathrm{i}\phi}+\sin^{2}\frac{\theta_{F}}{2}\mathrm{e}^{-\mathrm{i}\phi}\right),
\end{equation}
and the correlation function of the two-amplitude is 

\begin{equation}
\begin{aligned}\left\langle U_{{\bf k}{\bf k}'}^{++}U_{{\bf k}'{\bf k}}^{+-}\right\rangle = & \frac{1}{2\pi{\rm S}\rho_{F}\tau}\frac{\sin\theta_{F}}{1+\cos^{2}\theta_{F}}\times\Big[\cos\theta_{F}-\\
 & \cos\theta_{F}\left(\frac{\tau}{\tau_{0}}-\frac{\tau}{\tau_{z}}\right)\cos\left(\phi-\phi'\right)\\
 & -{\rm i}\left(\frac{\tau}{\tau_{0}}-\frac{\tau}{\tau_{z}}\right)\sin\left(\phi-\phi'\right)\Big].
\end{aligned}
\end{equation}

By leveraging the velocities, Green's functions, and the scattering
amplitude correlation functions, we derive the side-jump Hall conductivity
up to the fourth-order in disorder correlations. For instance, the
$\sigma_{xy}^{\mathrm{sj}1}$ can be evaluated as: 
\begin{equation}
\begin{aligned}\sigma_{xy}^{\mathrm{sj}1}= & \frac{e^{2}}{h}\int\frac{\mathrm{d}\phi}{2\pi}\frac{{\rm S}\mathrm{d}\phi'}{2\pi}V_{x{\bf k}}^{++}v_{y{\bf k}}^{-+}\left\langle U_{{\bf k}{\bf k}'}^{++}U_{{\bf k}'{\bf k}}^{+-}\right\rangle \\
 & \times\int\rho\left(\epsilon_{{\bf k}}\right)\mathrm{d}\epsilon_{{\bf k}}G_{+{\bf k}}^{R}G_{+{\bf k}}^{A}G_{-{\bf k}}^{R}\int\rho\left(\epsilon_{{\bf k}'}\right)\mathrm{d}\epsilon_{{\bf k}'}G_{+{\bf k}'}^{R}.
\end{aligned}
\end{equation}
With the assumption of the Ioffe-Regel condition $\epsilon_{F}\tau\gg1$,
and utilizing the residue theorem, we can perform the Green's function
calculations as 
\begin{gather}
\int\rho\left(\epsilon_{{\bf k}}\right)\mathrm{d}\epsilon_{{\bf k}}G_{+{\bf k}}^{R}G_{+{\bf k}}^{A}G_{-{\bf k}}^{R}\approx\frac{\pi\rho_{F}\tau}{\epsilon_{F}},\\
\int\rho\left(\epsilon_{{\bf k}'}\right)\mathrm{d}\epsilon_{{\bf k}'}G_{+{\bf k}'}^{R}=-\mathrm{i}\pi\rho_{F}.
\end{gather}
Meanwhile, the angular average of the velocities and scattering correlations
are 
\begin{equation}
\int\frac{\mathrm{d}\phi}{2\pi}\frac{{\rm S}\mathrm{d}\phi'}{2\pi}V_{x{\bf k}}^{++}v_{y{\bf k}}^{-+}\left\langle U_{{\bf k}{\bf k}'}^{++}U_{{\bf k}'{\bf k}}^{+-}\right\rangle =-\mathrm{i}\frac{\eta_{v}\lambda_{\parallel}^{2}}{4\pi\rho_{F}\tau}\frac{\sin^{2}\theta_{F}\cos\theta_{F}}{1+\cos^{2}\theta_{F}}.
\end{equation}
With the above equations, we found that 
\begin{equation}
\sigma_{xy}^{\mathrm{sj}1}=-\frac{e^{2}}{h}\frac{\mathfrak{b}}{1-\mathfrak{a}}\frac{\cos\theta_{F}}{4},
\end{equation}
where, for brevity, we introduce the shorthand notation \citep{LuHZ2013:PRB}
\begin{equation}
\mathfrak{a}=\frac{1}{2}\left(\frac{\tau}{\tau_{0}}-\frac{\tau}{\tau_{z}}\right)\frac{\sin^{2}\theta_{F}}{1+\cos^{2}\theta_{F}},\,\mathfrak{b}=\frac{1}{2}\frac{\sin^{2}\theta_{F}}{1+\cos^{2}\theta_{F}}.
\end{equation}
Analogously, the second side-jump diagrams can be evaluated as: 
\begin{equation}
\sigma_{xy}^{\mathrm{sj}2}=-\frac{e^{2}}{h}\frac{\mathfrak{a}}{1-\mathfrak{a}}\frac{\cos\theta_{F}}{4}.
\end{equation}
Taking into account symmetry-related diagrams, the total side-jump
Hall conductivity is
\begin{equation}
\sigma_{xy}^{\mathrm{sj}}=4\left(\sigma_{xy}^{\mathrm{sj}1}+\sigma_{xy}^{\mathrm{sj}2}\right)=-\frac{e^{2}}{h}\frac{\mathfrak{a}+\mathfrak{b}}{1-\mathfrak{a}}\cos\theta_{F}.
\end{equation}
The side-jump Hall conductivity contains an overall factor $\cos\theta_{F}$,
rendering it comparable in magnitude to the intrinsic anomalous Hall
conductance. Its dependence on disorder strength is implicit, entering
solely through the scattering-time combination $\frac{\tau}{\tau_{0}}-\frac{\tau}{\tau_{z}}$,
which captures the competition between nonmagnetic and spin-conserving
magnetic channels. Notably, spin-flipping impurities have no effect
on the side-jump contribution.

\subsection{Intrinsic and Extrinsic Skew-Scattering Mechanisms}

Having discussed the intricacies of side-jump scattering, we now turn
our focus to the skew-scattering Hall conductivity, where the dominant
contributions are illustrated in Fig. \ref{fig:AHE} (d1-d5). First,
for the intrinsic and extrinsic skew-scattering mechanisms denoted
by Fig. \ref{fig:AHE} (d1-d3), these diagrams can be evaluated as
\begin{equation}
\begin{aligned}\sigma_{xy}^{\mathrm{sk}1}= & \frac{e^{2}}{h}\int\frac{\mathrm{d}^{2}{\bf k}}{(2\pi)^{2}}\sum_{{\bf k}'{\bf k}''}V_{x{\bf k}}^{++}V_{y{\bf k}''}^{++}\left\langle U_{{\bf k}{\bf k}'}^{++}U_{{\bf k}'{\bf k}}^{+-}\right\rangle \\
 & \times\left\langle U_{{\bf k}'{\bf k}''}^{++}U_{{\bf k}''{\bf k}'}^{-+}\right\rangle G_{+{\bf k}}^{R}G_{+{\bf k}}^{A}G_{+{\bf k}'}^{R}G_{-{\bf k}'}^{A}G_{+{\bf k}''}^{R}G_{+{\bf k}''}^{A},
\end{aligned}
\end{equation}
\begin{equation}
\begin{aligned}\sigma_{xy}^{\mathrm{sk}2}= & \frac{e^{2}}{h}\int\frac{\mathrm{d}^{2}{\bf k}}{(2\pi)^{2}}\sum_{{\bf k}'{\bf k}''}V_{x{\bf k}}^{++}V_{y{\bf k}'}^{++}\left\langle U_{{\bf k}{\bf k}'}^{+-}U_{{\bf k}'{\bf k}}^{++}\right\rangle \\
 & \times\left\langle U_{{\bf k}'{\bf k}''}^{-+}U_{{\bf k}''{\bf k}'}^{++}\right\rangle G_{+{\bf k}}^{R}G_{+{\bf k}}^{A}G_{-{\bf k}'}^{R}G_{+{\bf k}'}^{R}G_{+{\bf k}'}^{A}G_{+{\bf k}''}^{R},
\end{aligned}
\end{equation}
\begin{equation}
\begin{aligned}\sigma_{xy}^{\mathrm{sk}3}= & \frac{e^{2}}{h}\int\frac{\mathrm{d}^{2}{\bf k}}{(2\pi)^{2}}\sum_{{\bf k}'{\bf k}''}V_{x{\bf k}}^{++}V_{y{\bf k}'}^{++}\left\langle U_{{\bf k}{\bf k}'}^{++}U_{{\bf k}'{\bf k}''}^{++}U_{{\bf k}''{\bf k}}^{++}\right\rangle \\
 & \times G_{+{\bf k}}^{R}G_{+{\bf k}}^{A}G_{+{\bf k}'}^{R}G_{+{\bf k}'}^{A}G_{+{\bf k}''}^{A}.
\end{aligned}
\end{equation}
In parallel with our earlier treatment of the side-jump Hall conductivity,
we can derive the expression (see Appendix \ref{app:A} for the definition
of $U_{0,z}^{3}$) by following a similar procedure:
\begin{equation}
\begin{aligned}\sigma_{xy}^{\mathrm{sk}1}= & -\frac{e^{2}}{h}\left(\frac{\mathfrak{a}}{1-\mathfrak{a}}\right)^{2}\frac{\cos\theta_{F}}{4},\end{aligned}
\end{equation}
\begin{equation}
\begin{aligned}\sigma_{xy}^{\mathrm{sk}2}= & -\frac{e^{2}}{h}\frac{\mathfrak{a}\mathfrak{b}}{\left(1-\mathfrak{a}\right)^{2}}\frac{\cos\theta_{F}}{4},\end{aligned}
\end{equation}
\begin{equation}
\begin{aligned}\sigma_{xy}^{\mathrm{sk}3}= & -\frac{e^{2}}{h}\frac{\epsilon_{F}\left(\pi\rho_{F}\tau\sin^{2}\theta_{F}\right)^{2}}{8\left(1-\mathfrak{a}\right)^{2}}\left(U_{3}^{0}\cos\theta_{F}-U_{3}^{z}\right).\end{aligned}
\end{equation}
As shown above, the skew-scattering Hall conductivity arising from
the diagrams in Fig. \ref{fig:AHE} (d1,d2) exhibits the same overall
factor $\cos\theta_{F}$ noted previously. In contrast, for the extrinsic
skew-scattering Hall conductivity arising from diagrams in Fig. \ref{fig:AHE}
(d3), this factor appears only for non-magnetic impurity scattering,
and is absent for the magnetic impurity aligned along the $z$-direction.

\subsection{Coherent Skew-Scattering Mechanism}

Additionally, there are two distinct classes of crossed diagrams that
contribute to the Hall conductivity through coherent skew-scattering
\citep{Ostrovsky2017:PRB,ZhangJX2024:PRB}, as illustrated in Fig.
\ref{fig:AHE}(d4,d5). The corresponding expressions for these Feynman
diagrams are: 
\begin{equation}
\begin{aligned}\sigma_{xy}^{\mathrm{sk}4}=\frac{e^{2}}{h}\int\frac{\mathrm{d}^{2}{\bf k}}{(2\pi)^{2}} & \sum_{{\bf k}',{\bf k}''}V_{x{\bf k}}^{++}V_{y,{\bf k}'}^{++}\left\langle U_{{\bf k}{\bf k}''}^{+-}U_{{\bf k}',{\bf k}+{\bf k}'-{\bf k}''}^{++}\right\rangle \\
 & \left\langle U_{{\bf k}''{\bf k}}^{-+}U_{{\bf k}+{\bf k}'-{\bf k}'',{\bf k}}^{++}\right\rangle G_{+{\bf k}}^{R}G_{-{\bf k}''}^{R}\\
 & G_{+,{\bf k}'}^{R}G_{+,{\bf k}'}^{A}G_{+,{\bf k}+{\bf k}'-{\bf k}''}^{A}G_{+{\bf k}}^{A},
\end{aligned}
\end{equation}
\begin{equation}
\begin{aligned}\sigma_{xy}^{\mathrm{sk}5}=\frac{e^{2}}{h}\int\frac{\mathrm{d}^{2}{\bf k}}{(2\pi)^{2}} & \sum_{{\bf k}',{\bf k}''}V_{x{\bf k}}^{++}V_{y{\bf k}'}^{++}\left\langle U_{{\bf k}{\bf k}''}^{+-}U_{{\bf k}''-{\bf k}+{\bf k}',{\bf k}'}^{++}\right\rangle \\
 & \left\langle U_{{\bf k}'',{\bf k}''-{\bf k}+{\bf k}'}^{-+}U_{{\bf k}'{\bf k}}^{++}\right\rangle G_{+{\bf k}}^{R}G_{-{\bf k}''}^{R}\\
 & G_{+,{\bf k}''-{\bf k}+{\bf k}'}^{R}G_{+,{\bf k}'}^{R}G_{+,{\bf k}'}^{A}G_{+{\bf k}}^{A}.
\end{aligned}
\end{equation}
In evaluating these diagrams, we observe that
\begin{equation}
\int\frac{\mathrm{d}^{2}{\bf k}''}{(2\pi)^{2}}G_{-{\bf k}''}^{R}G_{+,{\bf k}+{\bf k}'-{\bf k}''}^{A}\approx\int\frac{{\rm d}\phi''}{2\pi}\frac{-2\pi{\rm i}\rho_{F}}{2\epsilon_{F}+\frac{\partial\epsilon_{{\bf k}''}}{\partial{\bf k}''}\cdot\left({\bf k}+{\bf k}'\right)},
\end{equation}
\begin{equation}
\int\frac{\mathrm{d}^{2}{\bf k}''}{(2\pi)^{2}}G_{-{\bf k}''}^{R}G_{+,{\bf k}''-{\bf k}+{\bf k}'}^{R}\approx\int\frac{{\rm d}\phi''}{2\pi}\frac{-2\pi{\rm i}\rho_{F}}{2\epsilon_{F}+\frac{\partial\epsilon_{{\bf k}''}}{\partial{\bf k}''}\cdot\left({\bf k}-{\bf k}'\right)},
\end{equation}
thereby permitting their reduction to the form 
\begin{equation}
\mathrm{Re}\sigma_{xy}^{\mathrm{sk}4}=\frac{e^{2}}{h}\frac{\mathfrak{a}\left(g_{1}\mathfrak{a}+g_{2}\mathfrak{b}\right)}{\left(1-\mathfrak{a}\right)^{2}}\cos\theta_{F},
\end{equation}
\begin{equation}
\mathrm{Re}\sigma_{xy}^{\mathrm{sk}5}=\frac{e^{2}}{h}\frac{\mathfrak{a}\left(g_{3}\mathfrak{a}+g_{4}\mathfrak{b}\right)}{\left(1-\mathfrak{a}\right)^{2}}\cos\theta_{F},
\end{equation}
where the factors $g_{i}$ are 
\begin{equation}
\begin{cases}
g_{1}=2\,_{2}F_{1}\left(\frac{1}{2},\frac{3}{2};2;\sin^{4}\theta_{F}\right)-\frac{3}{2}\,_{2}F_{1}\left(\frac{1}{2},\frac{5}{2};3;\sin^{4}\theta_{F}\right),\\
g_{2}=\frac{2\left[2\,_{2}F_{1}\left(\frac{1}{2},\frac{1}{2};1;\sin^{4}\theta_{F}\right)-\,_{2}F_{1}\left(\frac{1}{2},\frac{3}{2};2;\sin^{4}\theta_{F}\right)-1\right]}{\sin^{2}\theta_{F}},\\
g_{3}=\frac{1}{2}g_{2}+\frac{1-2g_{1}}{\sin^{2}\theta_{F}},\\
g_{4}=-g_{1}-\frac{1}{2}g_{2},
\end{cases}
\end{equation}
and $\,_{2}F_{1}\left(a,b;c;z\right)$ is the Gauss hypergeometric
function. Both the X and $\Psi$ diagrams yield contributions proportional
to $\cos\theta_{F}$, rendering them comparable in magnitude to the
other scattering mechanisms.

\subsection{Summary}

In a culmination of our efforts, we assemble the Feynman diagrams
in Fig. \ref{fig:AHE}(d1-d5) to arrive at the final expression for
the skew-scattering Hall conductivity (in unit of $e^{2}/h$):

\begin{equation}
\begin{aligned}\sigma_{xy}^{\mathrm{sk}}= & 2\sigma_{xy}^{\mathrm{sk}1}+4\sigma_{xy}^{\mathrm{sk}2}+2\mathrm{Re}\sigma_{xy}^{\mathrm{sk}3}+2\mathrm{Re}\sigma_{xy}^{\mathrm{sk}4}+4\mathrm{Re}\sigma_{xy}^{\mathrm{sk}5}\\
= & \frac{\mathfrak{a}\left[\left(4g_{1}+8g_{3}-1\right)\mathfrak{a}-2\left(4g_{1}+1\right)\mathfrak{b}\right]}{2\left(1-\mathfrak{a}\right)^{2}}\cos\theta_{F}\\
 & -\frac{\epsilon_{F}\left(\pi\rho_{F}\tau\sin^{2}\theta_{F}\right)^{2}}{8\left(1-\mathfrak{a}\right)^{2}}\times\left(U_{3}^{0}\cos\theta_{F}-U_{3}^{z}\right).
\end{aligned}
\end{equation}

The extrinsic anomalous Hall conductivity has been obtained as $\sigma_{xy}^{\mathrm{ex}}=\sigma_{xy}^{\mathrm{sj}}+\sigma_{xy}^{\mathrm{sk}}$,
which can be divided as two parts. The side-jump, intrinsic and coherent
skew-scattering, and non-magnetic extrinsic skew-scattering contributions
to the Hall conductivity all share a common $\cos\theta_{F}$ factor.
Notably, in the scenario where the Fermi level resides within the
gapless regime characterized by $k_{F}<k_{c}$, this term is eliminated
due to the absence of a gap, as we have previously established in
Sec. \ref{sec:AHE}. The sole possibility for breaking the half-quantized
Hall conductivity arises from the skew scattering triggered by out-of-plane
magnetic impurities:
\begin{equation}
\sigma_{xy}^{\mathrm{ex}}=\frac{\epsilon_{F}\left(\pi\rho_{F}\tau\sin^{2}\theta_{F}\right)^{2}U_{3}^{z}}{8\left(1-\mathfrak{a}\right)^{2}}.
\end{equation}
Nevertheless, in our case, the third-order moment of disorder strength
vanishes, and thus $U_{3}^{z}=0$. As a consequence, these mechanisms
do not threaten the integrity of the half-quantized Hall conductivity.

\section{Discussion\label{sec:Conclusion}}

Regularizing a single gapless Dirac cone on a lattice while preserving
parity or time-reversal symmetry presents a fundamental obstruction,
as dictated by the fermion doubling theorem \citep{Nielsen1981:PLB,Shen2022:npjQM,Shen2026prb}.
By relaxing either time-reversal symmetry or locality condition, several
approaches to discretizing the massless Dirac fermion on a two-dimensional
lattice have been proposed \citep{SLAC1976:PRD,Wilson1974:PRD,Stacey1982:PRD,Beenakker2022:AP,Zou2023:PRB,Beenakker2023:AP}.
Among these, only the Wilson fermion formulation correctly reproduces
the half-quantized Hall conductance when the chemical potential lies
at the Dirac point. However, the Wilson fermion formulation is susceptible
to disorder-induced mass renormalization. Even weak disorder can open
a gap at the Dirac point and drive a transition into either a trivial
or topological insulating phase, thereby obscuring the half-quantized
Hall conductance.

Recent experimental advances have demonstrated that semi-magnetic
or tri-layer topological insulator heterostructures provide ideal
platforms for hosting a single massless Dirac fermion and observing
the half-quantized Hall effect \citep{Mogi2022:NatPhys,Zhuo-prl-2026,Yang-AdvMater-2026,Wang-prl2026}.
In these systems, the PAS phase emerges as the gapless Dirac cone
surface state that is protected by the bulk topological properties
of the topological insulator film. Notably, the PAS phase exhibits
remarkable robustness against scalar disorder scattering. The half-quantized
Hall conductance persists over a finite window of disorder strength
and chemical potential \citep{Bi2025:CommPhys,Bi2025:PRB}, demonstrating
significantly greater stability than lattice realizations based on
Wilson fermions. In the present work, we extend the previous analysis
to incorporate the effects of magnetic impurity scattering. This is
inherently relevant to the magnetic structure of topological insulator
films, since the magnetic dopants inevitably introduce structural
disorder. We have analytically calculated the contributions of side-jump
and skew-scattering mechanisms to the extrinsic anomalous Hall conductivity
in the PAS, up to fourth order in disorder strength. We find that
all such extrinsic contributions vanish, thereby preserving the half-quantization
of the Hall conductance and explaining the robustness of the half-quantized
Hall effect in the PAS.

\begin{figure}
\includegraphics[width=8.5cm]{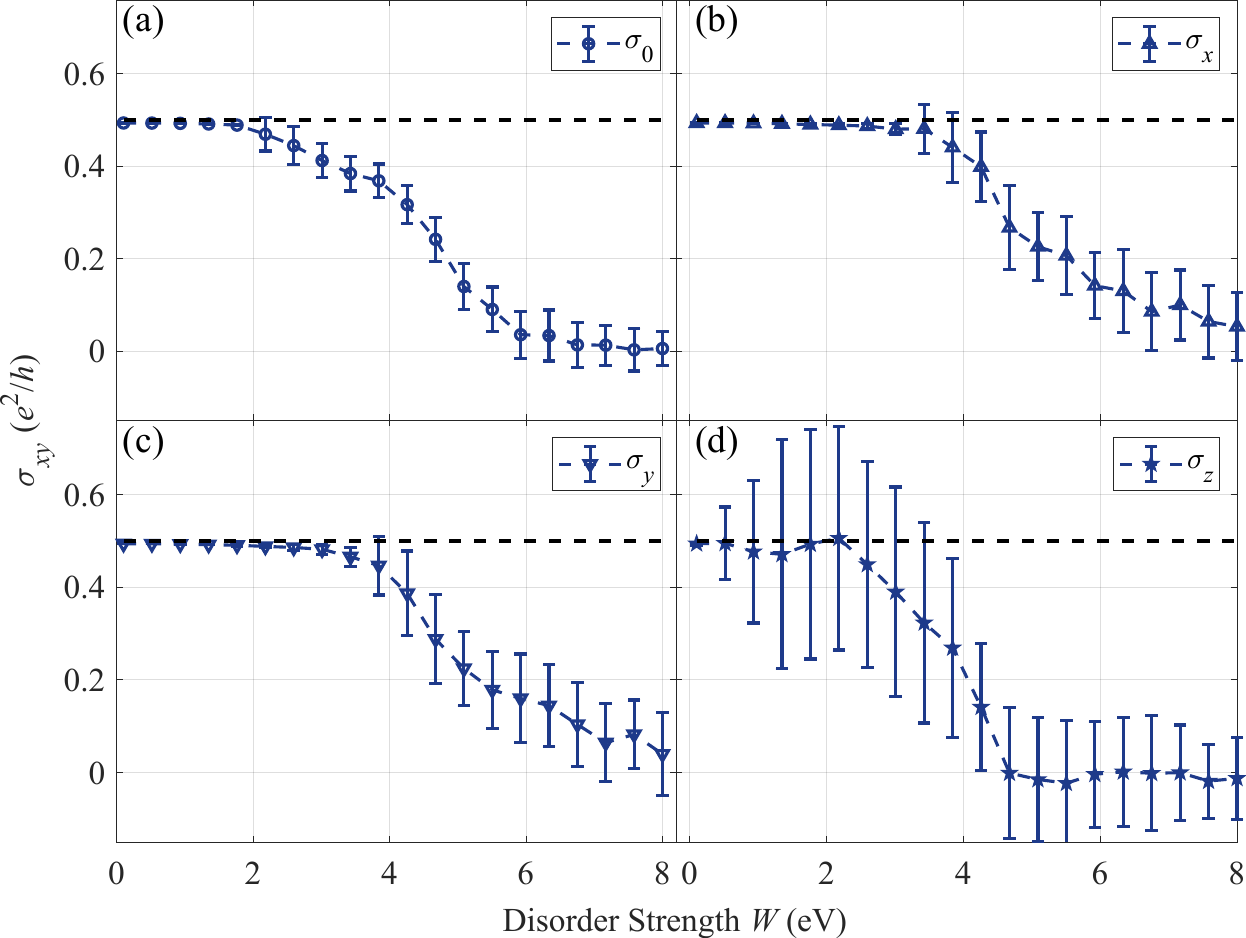}

\caption{Real-space Hall conductivity of the PAS calculated using Prodan's
formula \citep{Prodan2011:JPA,Bi2025:CommPhys,Bi2025:PRB}. The various
markers and legends correspond to different types of impurities, respectively.
System parameters: The simulations employed a system size of $L_{x}=L_{y}=28$
and $L_{z}=10$, with the top $3$ layers exhibiting ferromagnetic
order. Additional parameters \citep{Zhang2009:NatPhys} included $\lambda_{\parallel}=0.41$
eV, $\lambda_{\perp}=0.44$ eV, $t_{\parallel}=0.566$ eV, $t_{\perp}=0.40$
eV, $V_{0}=0.1$ eV, $m_{0}=0.28$ eV and $\epsilon_{F}=10$ meV.
Lattice constants were set to $a=b=1$ nm and $c=0.5$ nm \citep{Zhang2009:NatPhys}.
To minimize statistical fluctuations, 100 samples were averaged for
each data point.}

\label{fig:RSCN}
\end{figure}

To corroborate our analytical result, we perform numerical calculations
\citep{Prodan2011:JPA,Bi2025:CommPhys,Bi2025:PRB} of Hall conductance
on a real-space lattice based on the tight-binding model Eq.(\ref{eq:H-1}),
Eq. (\ref{eq:ZM}), and Eq. (\ref{eq:Uimp}) . In the presence of
scalar (Fig. (\ref{fig:RSCN}) (a)) or spin-flip impurities (Fig.
(\ref{fig:RSCN}) (b,c)), the system exhibits robust half-quantized
Hall plateaus with minimal fluctuations, in stark contrast to the
$\sigma_{z}$-type impurities, which yield only an approximate plateau
amidst substantial fluctuations. Furthermore, the plateau is even
more robust in the presence of spin-flip impurities than in the case
of scalar impurities. The observed half-quantized plateau persisting
in a finite range of disorder strength reinforces our analysis, ruling
out extrinsic contributions from side-jump and skew-scattering effects.

The validity of our results is governed by three key energy scales:
the Zeeman exchange field strength $V_{0}$, the bulk gap $m_{0}$
of topological insulator film, and the energy level broadening $\eta$
inside the gap induced by the Zeeman field. When the Fermi energy
lies within the gap of the massive Dirac cone, the intrinsic Hall
conductance becomes half-quantized, provided that $\eta\ll V_{0}$
\citep{Bi2025:CommPhys,Bi2025:PRB,fu2025nearly}. In our formulation,
this condition is naturally satisfied in the weak-scattering regime,
where $\eta$ is vanishingly small and disorder-induced corrections
to the Hall response from the massive Dirac cone are negligible. Another
subtle issue is about the effect of time-reversal symmetry breaking.
A genuine two-dimensional massless Dirac fermion possesses time-reversal
symmetry and exhibits weak antilocalization that evades the Anderson
localization. However, the parity anomaly arises from quantum corrections
that necessarily break parity and time-reversal symmetry, which appears
to undermine the symmetry protection that ensures delocalization.
Our analysis of scattering amplitude correlations in Sec. \ref{sec:Model}
(B) reveals that they are confined to the parity-invariant regime
$k<k_{c}$, and are strongly suppressed for $k>k_{c}$. It is precisely
in this outer region that the occupied states generate the half-quantized
Hall conductance. Since $k_{c}$ is directly tied to the bulk gap
$m_{0}$, we conclude that as long as the gap is not closed by disorder-induced
trivial states, the electronic states for $k>k_{c}$ remain undisturbed.
Consequently, weak antilocalization persists, protecting the PAS from
Anderson localization.

Experimentally, the minimal longitudinal conductivity near the Dirac
point has been measured to fall within the range of $0.5\sim0.67e^{2}/h$
\citep{Mogi2022:NatPhys,Zhuo-prl-2026,Yang-AdvMater-2026,Wang-prl2026},
thereby demonstrating its metallic transport character. However, due
to strong disorder scattering in the vicinity of the Dirac point ($\epsilon_{F}\tau\lesssim1$),
standard perturbation theory becomes uncontrolled, and the precise
value of the minimal conductivity remains inconclusive \citep{Zirnbauer1997:JMP,Zirnbauer2019:NPB,KLZ1992:PRB,LWK1993:PRL,WangZQ1996:PRL,Dolan1999:NPB,Tsvelik2007:PRB,Ohtsuki2005:PhysRep}.
The PAS provides a feasible framework for field-theoretic formulations
incorporating non-perturbative effect \citep{DTSon2015:PRX,Witten2016:AP,DavidTong2016:PRX,Chen2018:PRL,MaCT2018:AP,MaCT2018:FdP},
and may shed light on the critical theory associated with half-quantized
Hall effect. We leave these topics for future investigation.
\begin{acknowledgments}
This work was supported by the Research Grants Council, University
Grants Committee, Hong Kong under Grant No. 17301823; and Guangdong
Provincial Quantum Science Strategic initiative (GDZX230005); National
Natural Science Foundation of China (Grants No.12504049), Guangdong
Province Introduced Innovative R\&D Team Program (Grant No. 2023QN10X136),
Guangdong Basic and Applied Basic Research Foundation No. 2024A1515010430
and 2023A1515140008).
\end{acknowledgments}

\renewcommand\theequation{\Alph{section}\arabic{equation}}

\appendix

\section{Correlation Function of the Scattering Matrix with Three Amplitudes\label{app:A}}

In the Feynman diagram for the skew scattering, the three-amplitude
correlation function $\left\langle U_{\alpha,{\bf kk}'}^{SS}U_{\alpha,{\bf k}'{\bf k}''}^{SS}U_{\alpha,{\bf k}''{\bf k}}^{SS}\right\rangle $
is involved. Introducing the summation 
\begin{equation}
\begin{array}{rl}
U_{0/z}^{3,R}= & {\displaystyle \frac{1}{{\rm S}}\sum_{i_{z},{\bf r}_{j}}\sum_{i_{z}^{\prime},{\bf r}_{j}^{\prime}}\sum_{i_{z}^{\prime\prime},{\bf r}_{j}^{\prime\prime}}}\left\langle u_{0/z,i_{z},{\bf r}_{j}}u_{0/z,i_{z}^{\prime},{\bf r}_{j}^{\prime}}u_{0/z,i_{z}^{\prime\prime},{\bf r}_{j}^{\prime\prime}}\right\rangle \\
 & \mathrm{e}^{-\mathrm{i}\left({\bf k}-{\bf k}'\right)\cdot{\bf r}_{j}}\mathrm{e}^{-\mathrm{i}\left({\bf k}'-{\bf k}''\right)\cdot{\bf r}_{j}^{\prime}}\mathrm{e}^{-\mathrm{i}\left({\bf k}''-{\bf k}\right)\cdot{\bf r}_{j}^{\prime\prime}}\\
 & \left(\delta_{1,{\bf k}{\bf k}'}\cos\frac{\vartheta_{{\bf k}}-\vartheta_{{\bf k}'}}{2}+\mathrm{i}\delta_{2,{\bf k}{\bf k}'}\sin\frac{\vartheta_{{\bf k}}+\vartheta_{{\bf k}'}}{2}\right)\\
 & \left(\delta_{1,{\bf k}'{\bf k}''}\cos\frac{\vartheta_{{\bf k}'}-\vartheta_{{\bf k}''}}{2}+\mathrm{i}\delta_{2,{\bf k}'{\bf k}''}\sin\frac{\vartheta_{{\bf k}'}+\vartheta_{{\bf k}''}}{2}\right)\\
 & \left(\delta_{1,{\bf k}''{\bf k}}\cos\frac{\vartheta_{{\bf k}''}-\vartheta_{{\bf k}}}{2}+\mathrm{i}\delta_{2,{\bf k}''{\bf k}}\sin\frac{\vartheta_{{\bf k}''}+\vartheta_{{\bf k}}}{2}\right),
\end{array}
\end{equation}
\begin{equation}
\begin{array}{rl}
U_{x/y}^{3,R}= & {\displaystyle \frac{1}{{\rm S}}\sum_{i_{z},{\bf r}_{j}}\sum_{i_{z}^{\prime},{\bf r}_{j}^{\prime}}\sum_{i_{z}^{\prime\prime},{\bf r}_{j}^{\prime\prime}}}\left\langle u_{x/y,i_{z},{\bf r}_{j}}u_{x/y,i_{z}^{\prime},{\bf r}_{j}^{\prime}}u_{x/y,i_{z}^{\prime\prime},{\bf r}_{j}^{\prime\prime}}\right\rangle \\
 & \mathrm{e}^{-\mathrm{i}\left({\bf k}-{\bf k}'\right)\cdot{\bf r}_{j}}\mathrm{e}^{-\mathrm{i}\left({\bf k}'-{\bf k}''\right)\cdot{\bf r}_{j}^{\prime}}\mathrm{e}^{-\mathrm{i}\left({\bf k}''-{\bf k}\right)\cdot{\bf r}_{j}^{\prime\prime}}\\
 & \left(R\delta_{3,{\bf k}{\bf k}'}\cos\frac{\vartheta_{{\bf k}}+\vartheta_{{\bf k}'}}{2}+\mathrm{i}\delta_{4,{\bf k}{\bf k}'}\sin\frac{\vartheta_{{\bf k}}-\vartheta_{{\bf k}'}}{2}\right)\\
 & \left(R\delta_{3,{\bf k}'{\bf k}''}\cos\frac{\vartheta_{{\bf k}'}+\vartheta_{{\bf k}''}}{2}+\mathrm{i}\delta_{4,{\bf k}'{\bf k}''}\sin\frac{\vartheta_{{\bf k}'}-\vartheta_{{\bf k}''}}{2}\right)\\
 & \left(R\delta_{3,{\bf k}''{\bf k}}\cos\frac{\vartheta_{{\bf k}''}+\vartheta_{{\bf k}}}{2}+\mathrm{i}\delta_{4,{\bf k}''{\bf k}}\sin\frac{\vartheta_{{\bf k}''}-\vartheta_{{\bf k}}}{2}\right),
\end{array}
\end{equation}
where the factors $\delta_{1,{\bf k}{\bf k}'}=\varphi_{{\bf k}}^{\dag}\left(+,l_{z}\right)\varphi_{{\bf k}'}\left(+,l_{z}\right)$,
$\delta_{2,{\bf k}{\bf k}'}=\varphi_{{\bf k}}^{\dag}\left(+,l_{z}\right)\chi_{{\bf k}'}\left(+,l_{z}\right)$,
$\delta_{3,{\bf k}{\bf k}'}=\varphi_{{\bf k}}^{\dag}\left(+,l_{z}\right)\chi_{{\bf k}'}\left(-,l_{z}\right)$,
and $\delta_{4,{\bf k}{\bf k}'}=\varphi_{{\bf k}}^{\dag}\left(+,l_{z}\right)\varphi_{{\bf k}'}\left(-,l_{z}\right)$.
Then the three-amplitude correlation function takes on a compact form
\begin{equation}
\left\langle U_{\alpha,{\bf kk}'}^{RR}U_{\alpha,{\bf k}'{\bf k}''}^{RR}U_{\alpha,{\bf k}''{\bf k}}^{RR}\right\rangle =\frac{U_{\alpha}^{3,R}}{{\rm S}^{2}}\sigma_{\alpha}\otimes\sigma_{\alpha}\otimes\sigma_{\alpha}.
\end{equation}
For the short-ranged Anderson impurity potential considered in the
present work, the third-order moment of the random variable vanishes,
\begin{equation}
\left\langle u_{\alpha,i_{z},{\bf r}_{j}}u_{\alpha,i_{z}^{\prime},{\bf r}_{j}^{\prime}}u_{\alpha,i_{z}^{\prime\prime},{\bf r}_{j}^{\prime\prime}}\right\rangle =0.
\end{equation}
Therefore, the representative diagram in Fig. (\ref{fig:AHE})(d3)
has no contribution.

\section{Proof of Vanishing $\gamma_{3,4}$\label{app:B}}

As demonstrated in Ref. \citep{KZ2024:SciPost}, the solution form
of $\xi_{1,2}$ are slightly different for $m_{0}\left({\bf k}\right)t_{\perp}/\lambda_{\perp}^{2}>1/4$
and $m_{0}\left({\bf k}\right)t_{\perp}/\lambda_{\perp}^{2}<1/4$.
For the case $m_{0}\left({\bf k}\right)t_{\perp}/\lambda_{\perp}^{2}>1/4$,
both $\xi_{1,2}$ are complex, and we have $\xi_{1}=\xi_{2}^{*}$.
The finite-size gap decays exponentially when $L_{z}$ increases.
While for $m_{0}\left({\bf k}\right)t_{\perp}/\lambda_{\perp}^{2}<1/4$,
both $\xi_{1,2}$ are purely imaginary. The finite-size gap decays
exponentially yet slower than the former case. In the following, we
prove that the $\gamma_{3,{\bf kk}'}$ and $\gamma_{4,{\bf kk}'}$
vanish identically in the gapless regime under consideration.

\textbf{Case} \textbf{I} (complex $\xi_{1}=\xi_{2}^{*}$ and $\xi_{1}^{\prime}=\xi_{2}^{\prime*}$):
we can choose $\mathrm{Im}\left(\xi_{1}\right)>0$, and thus $\tan\xi_{1}l\simeq\mathrm{i}$,
$\tan\xi_{2}l\simeq-\mathrm{i}$ for large $l$. Besides, $\cos\xi_{1}=\left(\cos\xi_{2}\right)^{*}$.
Moreover, the following equation 
\begin{equation}
\left(M+2t_{\perp}\cos\xi\right)^{2}+\lambda_{\perp}^{2}\sin^{2}\xi=0
\end{equation}
implies that the solution takes the form 
\begin{equation}
M+2t_{\perp}\cos\xi_{\alpha}=\left(-1\right)^{\alpha}\mathrm{i}\lambda_{\perp}\sin\xi_{\alpha}.
\end{equation}
Therefore, in the parity-invariant regime we have
\begin{equation}
\eta_{{\bf k},1}\simeq\frac{2\left(\cos\xi_{1}-\cos\xi_{2}\right)}{\mathrm{i}\left(\sin\xi_{1}+\sin\xi_{2}\right)}=-\frac{\lambda_{\perp}}{t_{\perp}}.
\end{equation}
The next thing we need to prove is 
\begin{equation}
\mathrm{Im}\left(\frac{\cos\xi_{1}l_{z}}{\cos\xi_{1}l}\right)=\mathrm{sgn}\left(l_{z}\right)\mathrm{Im}\left(\frac{\sin\xi_{1}l_{z}}{\sin\xi_{1}l}\right)
\end{equation}
To see this, for $l_{z}=l$ we find the expression holds naturally.
Then for $l_{z}=l-1$ we have 
\begin{equation}
\frac{\cos\xi_{1}\left(l-1\right)}{\cos\xi_{1}l}\simeq\cos\xi_{1}+\mathrm{i}\sin\xi_{1},
\end{equation}
\begin{equation}
\frac{\sin\xi_{1}\left(l-1\right)}{\sin\xi_{1}l}\simeq\cos\xi_{1}+\mathrm{i}\sin\xi_{1},
\end{equation}
and repeat this procedure we can prove the identity for $l_{z}>0$.
And the proof for $l_{z}=-l$ side is similar except for the extra
minus sign. Based on the above results,, we find that 
\begin{equation}
\gamma_{3}=\left|C_{{\bf k}}C_{{\bf k}'}\right|^{2}{\displaystyle \sum_{i_{z}}}\lambda_{\perp}^{4}\left|f_{{\bf k},+}^{*}f_{{\bf k}',-}-f_{{\bf k},-}^{*}f_{{\bf k}',+}\right|^{2}=0,
\end{equation}
\begin{equation}
\gamma_{4}=\left|C_{{\bf k}}C_{{\bf k}'}\right|^{2}{\displaystyle \sum_{i_{z}}}\lambda_{\perp}^{4}\left|f_{{\bf k},+}^{*}f_{{\bf k}',+}-f_{{\bf k},-}^{*}f_{{\bf k}',-}\right|^{2}=0.
\end{equation}

\textbf{Case} \textbf{II} (purely imaginary $\xi_{1,2}$ and $\xi_{1,2}^{\prime}$):
we set $\xi_{1}=\mathrm{i}\zeta_{1}$ and $\xi_{2}=\mathrm{i}\zeta_{2}$,
with $\zeta_{1,2}>0$. Now we have $\tan\xi_{1,2}l=\mathrm{i}\tanh\zeta_{1,2}l\simeq\mathrm{i}$
and 
\begin{equation}
M+2t_{\perp}\cos\xi_{1,2}=-\mathrm{i}\lambda_{\perp}\sin\xi_{1,2}.
\end{equation}
Therefore, 
\begin{equation}
\eta_{{\bf k},1}\simeq\frac{-2\left(\cos\xi_{1}-\cos\xi_{2}\right)}{-\mathrm{i}\left(\sin\xi_{1}-\sin\xi_{2}\right)}=-\frac{\lambda_{\perp}}{t_{\perp}},
\end{equation}
Again, we find that $f_{+}\left(l\right)=f_{-}\left(l\right)=0$ holds.
Reducing into the $l_{z}=l-1$ layer we find that 
\begin{equation}
f_{+}\left(l-1\right)=f_{-}\left(l-1\right)\simeq\cosh\zeta_{1}-\sinh\zeta_{1}-\cosh\zeta_{2}+\sinh\zeta_{2}.
\end{equation}
Following the same derivation as in the previous case, we conclude
that 
\begin{equation}
f_{{\bf k},+}\left(l_{z}\right)=\mathrm{sgn}\left(l_{z}\right)f_{{\bf k},-}\left(l_{z}\right)
\end{equation}
which are purely real functions. Therefore, the wavefunction shape
factors simplify to 
\begin{equation}
\gamma_{3}=\left|C_{{\bf k}}C_{{\bf k}'}\right|^{2}{\displaystyle \sum_{i_{z}}}\lambda_{\perp}^{4}\left|f_{{\bf k},+}f_{{\bf k}',-}-f_{{\bf k},-}f_{{\bf k}',+}\right|^{2}=0,
\end{equation}
\begin{equation}
\gamma_{4}=\left|C_{{\bf k}}C_{{\bf k}'}\right|^{2}{\displaystyle \sum_{i_{z}}}\lambda_{\perp}^{4}\left|f_{{\bf k},+}f_{{\bf k}',+}-f_{{\bf k},-}f_{{\bf k}',-}\right|^{2}=0.
\end{equation}

\textbf{Case} \textbf{III} (complex $\xi_{1,2}$ and purely imaginary
$\xi_{1,2}^{\prime}$): from the above analysis, we see that $\eta_{{\bf k},1}=\eta_{{\bf k}',1}=-\lambda_{\perp}/t_{\perp}$,
and 
\begin{equation}
\begin{aligned}f_{{\bf k},+}\left(l_{z}\right) & =\mathrm{sgn}\left(l_{z}\right)f_{{\bf k},-}\left(l_{z}\right)\end{aligned}
\end{equation}
Consequently, we conclude that $\gamma_{3,{\bf kk}'}=\gamma_{4,{\bf kk}'}=0$
holds identically throughout the surface state regime.

In Fig. \ref{fig:correlation}, we present the numerical calculation
of $\gamma_{3,{\bf kk}'}$ and $\gamma_{4,{\bf kk}'}$. The red boxes
denote the boundary where $m_{0}({\bf k})=m_{0}({\bf k}^{\prime})=0$.
As we can see, inside the red boxes, both $\gamma_{3,{\bf kk}'}$
and $\gamma_{4,{\bf kk}'}$ soon converges to zero. The larger $L_{z}$
becomes, the more rapidly they disappear.

%

\end{document}